\documentclass[journal]{IEEEtran}
\usepackage{times}
\usepackage{epsfig}
\usepackage{graphicx}
\usepackage{amsmath}
\usepackage{amssymb}
\usepackage{multirow}
\usepackage{cite}
\usepackage{array}
\usepackage{url}
\usepackage{booktabs}
\usepackage{mathtools}
\usepackage{makecell}
\usepackage{mathrsfs}
\usepackage[dvipsnames]{xcolor}
\usepackage{soul}
\usepackage{bbm}
\usepackage{amsthm}
\usepackage{subfigure}
\usepackage{float}
\usepackage{marvosym}
\definecolor{bananayellow}{rgb}{1.0, 0.88, 0.21}
\definecolor{chromeyellow}{rgb}{1.0, 0.65, 0.0}

\ifCLASSINFOpdf
\else
\fi

\begin{document}

\indent

© 2024 IEEE. Personal use of this material is permitted. Permission from IEEE must be obtained for all other uses, in any current or future media, including reprinting/republishing this material for advertising or promotional purposes, creating new collective works, for resale or redistribution to servers or lists, or reuse of any copyrighted component of this work in other works. Digital Object Identifier 10.1109/TIP.2024.3515873

%
\title{Modeling Dual-Exposure Quad-Bayer Patterns for Joint Denoising and Deblurring}
%
%
%

\author{Yuzhi~Zhao,~\IEEEmembership{Graduate~Student~Member,~IEEE,}
        Lai-Man~Po,~\IEEEmembership{Senior~Member,~IEEE,}
        \\ Xin~Ye, Yongzhe~Xu, Qiong~Yan
\thanks{Manuscript received October 14, 2022, revised September 21, 2023, accepted December 3, 2024. \textit{(Corresponding author: Yuzhi Zhao)}.}
\thanks{Y. Zhao, L.-M. Po are with the Department of Electronic Engineering, City University of Hong Kong, Hong Kong, China (e-mail: yzzhao2-c@my.cityu.edu.hk; eelmpo@cityu.edu.hk).}
\thanks{X. Ye, Y. Xu, Q. Yan are with the SenseTime Research and Tetras.AI, China (e-mail: yexinzju@gmail.com; xuyongzhe94@gmail.com; sophie.yanqiong@gmail.com).}
}

\markboth{IEEE Transactions on Image Processing}%
{Shell \MakeLowercase{\textit{Zhao et al.}}: Joint Image Deblurring and Denoising with Dual-exposure Quad-Bayer Sensors}
%

\maketitle

\begin{abstract}

Image degradation caused by noise and blur remains a persistent challenge in imaging systems, stemming from limitations in both hardware and methodology. Single-image solutions face an inherent tradeoff between noise reduction and motion blur. While short exposures can capture clear motion, they suffer from noise amplification. Long exposures reduce noise but introduce blur. Learning-based single-image enhancers tend to be over-smooth due to the limited information. Multi-image solutions using burst mode avoid this tradeoff by capturing more spatial-temporal information but often struggle with misalignment from camera/scene motion. To address these limitations, we propose a physical-model-based image restoration approach leveraging a novel dual-exposure Quad-Bayer pattern sensor. By capturing pairs of short and long exposures at the same starting point but with varying durations, this method integrates complementary noise-blur information within a single image. We further introduce a Quad-Bayer synthesis method (B2QB) to simulate sensor data from Bayer patterns to facilitate training. Based on this dual-exposure sensor model, we design a hierarchical convolutional neural network called QRNet to recover high-quality RGB images. The network incorporates input enhancement blocks and multi-level feature extraction to improve restoration quality. Experiments demonstrate superior performance over state-of-the-art deblurring and denoising methods on both synthetic and real-world datasets. The code, model, and datasets are publicly available at \url{https://github.com/zhaoyuzhi/QRNet}.

\end{abstract}


%
\IEEEpeerreviewmaketitle

\section{Introduction}

\IEEEPARstart{M}{ODERN} imaging pipelines inherently suffer from noise and blur artifacts stemming from fundamental hardware constraints, including shot noise, read noise, limited sensor resolution, quantization error, and suboptimal capture conditions such as low light. Additionally, motion blur remains problematic, particularly with handheld cameras. Noise and blur reduction have thus represented long-standing challenges in computational photography. With the proliferation of digital photography, producing high-quality images under real-world conditions is an important need. The straightforward approach of extending exposures gathers more photons but requires static scenes, often impractical outside controlled settings. Effective post-processing methods for noise and blur are thus essential.

Denoising typically relies on accurate noise models, often trained on RGB data despite noise arising in raw sensor data. Recent works have aimed to simulate raw noise more effectively \cite{brooks2019unprocessing, zamir2020cycleisp} or improve noise model estimation \cite{abdelhamed2019noise, wei2020physics, wang2020practical, zhang2021rethinking, maleky2022noise}, but texture preservation remains lacking. For deblurring, the common blind setting of unknown kernel or motion complicates modeling, while domain gaps between synthetic and real blur persist \cite{kupyn2018deblurgan, kupyn2019deblurgan, cho2021rethinking}. Joint denoising-deblurring approaches \cite{mustaniemi2018lsd, chang2021low, zhao2022d2hnet} using short- and long-exposure burst images provide more signal, but unaligned inputs make fusion difficult.

\begin{figure}[t]
\centering
\includegraphics[width=0.95\linewidth]{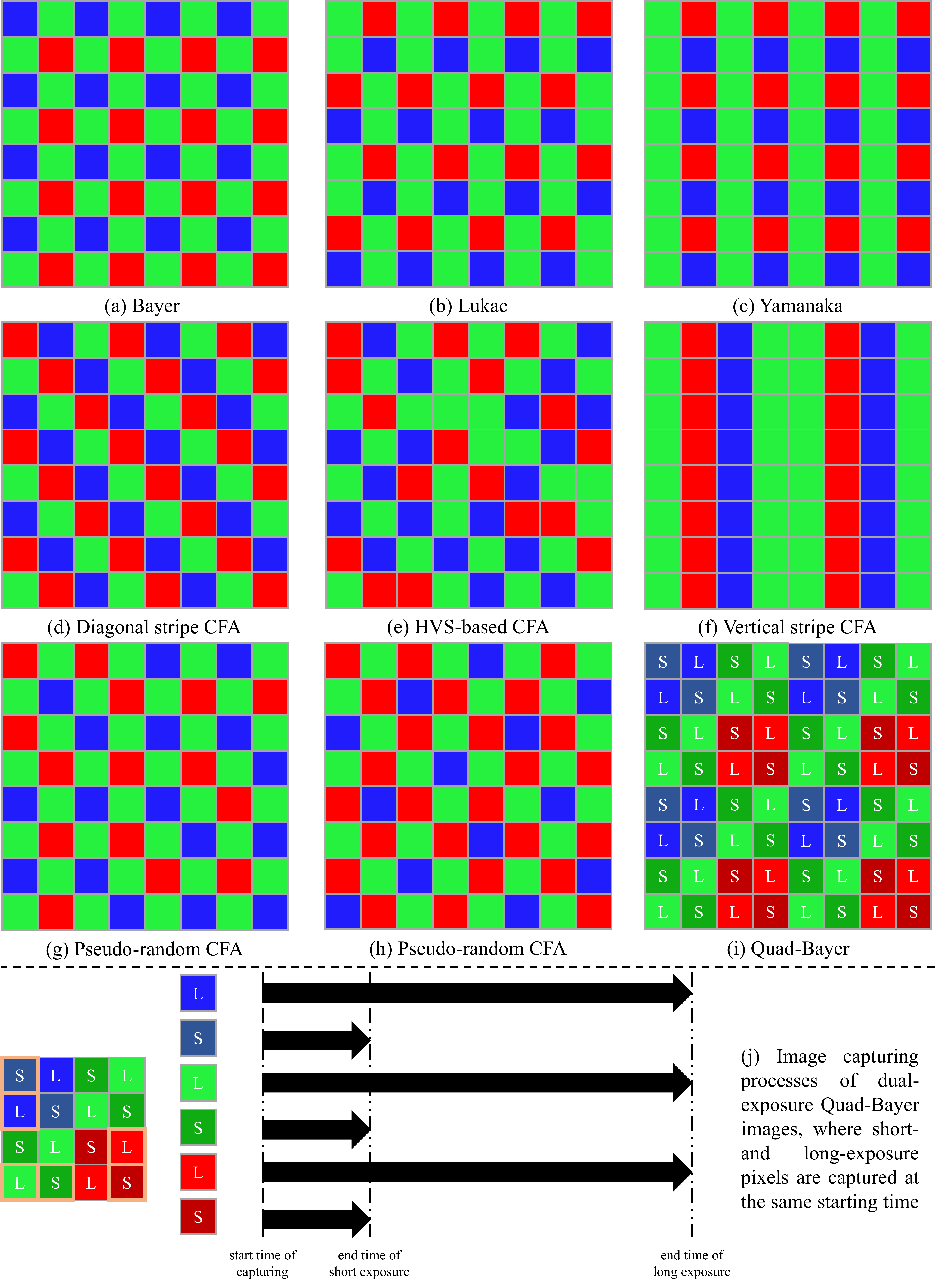}
\centering
\caption{Illustration of (a) Normal Bayer pattern, (b-h) Other Bayer patterns with single exposure, (i) Quad-Bayer pattern with both long and short exposures, and (j) Image capturing process of dual-exposure Quad-Bayer images.}
\label{cfa}
\end{figure}

The characteristics of short- and long-exposure images exhibit a complementary nature, with short-exposure pixels capturing sharp edges and long-exposure pixels containing minimal noise and accurate colors. By merging these two exposures, we can effectively restore textures while accurately estimating blur kernels simultaneously. A recent advancement in the field of image processing is the introduction of a sensor known as Quad-Bayer \cite{kim2021under, jiang2021hdr}, which incorporates two types of exposures within a single mosaic pattern. This pattern, serving as the minimum unit of the color filter array (CFA), possesses a resolution of 4$\times$4, as depicted in Figure \ref{cfa} (i). To fully leverage the capabilities of the Quad-Bayer sensor, we propose the development of a joint pipeline for image deblurring and denoising. Our proposed solution offers two significant advantages:

\begin{enumerate}\setlength{\itemsep}{-0.0cm}

\item The utilization of pixels from both short and long exposures simultaneously enhances the image reconstruction task by incorporating a larger amount of information compared to single-image methods \cite{brooks2019unprocessing, zamir2020cycleisp, abdelhamed2019noise, wei2020physics, wang2020practical, zhang2021rethinking, kupyn2018deblurgan, kupyn2019deblurgan, cho2021rethinking}.

\item The synchronization of pixels from both exposure types, as illustrated in Figure \ref{cfa} (j), eliminates any misalignment concerns typically encountered in burst-image methods \cite{mustaniemi2018lsd, chang2021low, zhao2022d2hnet}.

\end{enumerate}

We propose QRNet, an end-to-end neural network for joint deblurring and denoising of Quad-Bayer RAW images. To accommodate the Quad-Bayer layout, we employ the Pixel Unshuffle technique \cite{shi2016real} to reorganize the colors and exposures for the single-channel input, resulting in 16 channels. To mitigate the presence of discontinuous artifacts caused by downsampling, we introduce an input augmentation block (IEB). This block incorporates three additional branches, obtained through average pooling and convolutions, alongside the 16 channels obtained via Pixel Shuffle. Subsequently, all branches are merged using a channel attention scheme. Furthermore, QRNet is structured as a multi-level architecture, facilitating feature interactions between levels and enhancing the network's ability to represent features effectively.

To facilitate the training and evaluation of our proposed imaging pipeline, we have introduced a new synthetic dataset called the Quad-Bayer to RGB dataset (QR dataset). This dataset is essential as it is challenging to amass a large collection of Quad-Bayer and RGB image pairs using a camera equipped with a Quad-Bayer sensor that allows for flexible adjustment of camera parameters such as f-number, ISO, and focal length. The images in the dataset were originally captured using a DSLR camera (Nikon D5200) and encompass diverse scenes, including urban environments, parks, forests, mountains, and more. To generate the synthetic Quad-Bayer and RGB image pairs, we sample Quad-Bayer counterparts from pairs of long- and short-exposure Bayer RAW images. To ensure proper alignment between the synthetic Quad-Bayer and RGB image pairs, we propose a Bayer to Quad-Bayer (B2QB) sampling scheme. The resulting dataset comprises 701 training pairs and 30 validation pairs, all with a resolution of 3016$\times$2008 pixels.

We conducted a comprehensive evaluation of QRNet using both the QR dataset and real Quad-Bayer data. Both qualitative and quantitative experiments have shown that the proposed pipeline is capable of restoring images with intense noises or extremely blurry scenes, which outperforms existing approaches. Overall, the main contributions of this paper are as follows:

\begin{enumerate}\setlength{\itemsep}{-0.0cm}

\item We propose a novel solution that tackles the challenges of denoising and deblurring using Quad-Bayer sensors in a joint manner;

\item We introduce a Bayer to Quad-Bayer sampling method (B2QB) and curate a substantial QR dataset comprising 731 aligned Quad-Bayer and RGB image pairs;

\item We conduct extensive experiments, comparing image denoising and deblurring methods in both RGB space and RAW space, to demonstrate the significant advancements achieved by QRNet.

\end{enumerate}

In the following, we will review some more background and previous works in Section II, then introduce our method and the QR dataset in Section III and IV respectively, followed by our detailed experiment and analysis in Section V. Finally, we conclude the paper in Section VI.

\section{Related Work}

\noindent \textbf{Color Filter Array (CFA) and Quad-Bayer Pattern.} Image sensors are widely applied in modern color digital cameras. Considering the restriction on camera sizes, especially for mobile phone cameras, portable image sensors are in great demand. Color filters, as a significant component of image sensors, are designed to reduce the size and cost. These color filters are arranged in a mosaic pattern and each filter allows only one of the primary colors (R, G, B) to pass through. The arrangement is called a color filter array (CFA). For each pixel, only one color channel, either R, G, or B, is recorded. The missing ones are reconstructed by demosaicking. The first CFA was proposed by Bayer \cite{bayer1976color}, as shown in Figure \ref{cfa} (a). The minimum unit of the Bayer pattern is a 2$\times$2 sub-block, where the green channel is sampled at every other pixel and red/blue channels are sampled once. Later on, many other sampling configurations have been proposed such as Lukac CFA \cite{lukac2005color}, Yamanaka CFA \cite{yamanaka1977solid}, and HVS-based CFA \cite{parmar2004perceptually}. They are shown in Figure \ref{cfa} (b), (c), and (e), respectively. Though some of them improve the reconstruction accuracy or minimize an error, the sensor only records pixels under a single exposure. Recently, a novel sensor pixel layout called Quad-Bayer was introduced by Sony. The Quad-Bayer filter is an extension of the classic Bayer pattern, as shown in Figure \ref{cfa} (i). The structure enables two exposures within a group of four pixels thus avoiding multi-frame photography. It also alleviates the ghosting issue compared with classical multi-frame photography.

\noindent \textbf{Image Deblurring.} The target of image deblurring is estimating the blur kernel in the form of a convolution kernel or a motion field and then reconstructing the clear image. Deblurring methods can be categorized into blind and non-blind (i.e., blur convolution kernel or a motion field is known). Most of the classical methods addressed the non-blind deblurring issue by assuming blur kernel statistics and performing deconvolution. For instance, the Lucy-Richardson algorithm \cite{richardson1972bayesian} assumed a point spread function and the observed image should follow a Poisson distribution. However, it suffers from ringing artifacts. To address that issue, some methods used additional priors as regularizations such as GMM \cite{fergus2006removing}, patch \cite{sun2013edge}, sparsity \cite{krishnan2011blind}, and \emph{$l_0$} regularization \cite{pan2014deblurring}.

Since neural networks can extract rich data prior from large numbers of training pairs, they have been widely used to address the non-blind deblurring problem. From the perspective of network architecture, Sun \emph{et al.} \cite{sun2015learning} first proposed a multi-layer network to estimate blur kernel. Gong \emph{et al.} \cite{gong2017motion} utilized a fully convolutional network to produce a pixel-wise motion flow map. The output images are obtained by performing deconvolution for both of them. Nah \emph{et al.} \cite{nah2017deep} proposed an end-to-end non-blind deblurring method, which has three levels with different input resolutions. It was enhanced by SRN \cite{tao2018scale}, DMPHN \cite{zhang2019deep}, and MPRNet \cite{zamir2021multi} through a similar multi-scale training strategy. Furthermore, recent works have combined other generative models such as RNN-Deblur \cite{zhang2018dynamic}, Flow-guided Deblur \cite{yuan2020efficient}, Multi-stage Deblurring \cite{park2020multi}, and Diffusion-guided Deblurring \cite{chen2023hierarchical}. From the perspective of training strategy, DeblurGAN \cite{kupyn2018deblurgan} and DeblurGANv2 \cite{kupyn2019deblurgan} introduced adversarial training \cite{goodfellow2014generative} to existing architectures. More recently, some works have developed self-supervised training strategies such as Reblur2Deblur \cite{chen2018reblur2deblur}, Class-Specific Deblurring \cite{nimisha2018unsupervised}, Deblurring by Reblurring \cite{zhang2020deblurring}, SelfDeblur \cite{ren2020neural}, and Meta-Auxiliary Learning \cite{chi2021test}.

\noindent \textbf{Image Denoising.} The image denoising aims to remove noises and artifacts in an image. It also can be categorized into blind and non-blind denoising (i.e., noise model or noise kernel is known). Traditional non-blind image denoising methods often assumed a noise model first and then designed effective filters, e.g., diffusion \cite{perona1990scale}, total variation \cite{rudin1992nonlinear}, wavelet coring \cite{simoncelli1996noise}, non-local means \cite{buades2005non}, Wiener filter \cite{chen2006new}, sparse coding \cite{aharon2006k, elad2006image, mairal2009non} block-matching and 3-D filtering (BM3D) \cite{dabov2007image}. The noise model was often assumed to be additive white Gaussian noise (AWGN) or Poisson-Gaussian model (G-P). Though these approaches achieved good results on such specific noise patterns, they are not robust to real noises and have relatively high computational complexity.

To address the blind denoising problem, recent works often used neural networks including \cite{xie2012image, mao2016image, chen2016trainable, zhang2017learning, zhang2017beyond, tai2017memnet, zhang2018ffdnet, liu2018multi, anwar2019real, tian2020attention, liu2020densely, zamir2020learning, cheng2021nbnet, zamir2021multi}. Due to the strong fitting capability of neural networks, blind denoising can be conducted during inference following extensive training on real-world noisy-clean image pairs. As for the network design, researches have considered U-Net \cite{ronneberger2015u}, residual blocks \cite{he2016deep}, attention blocks \cite{anwar2019real}, flow model \cite{guo2024toward}, and diffusion model \cite{zhu2023denoising}. Furthermore, to estimate the noise level for training, CBDNet \cite{guo2019toward}, VDN \cite{yue2019variational}, and AINDNet \cite{kim2020transfer} combine a denoising network and a noise estimation network. Some works \cite{lehtinen2018noise2noise, batson2019noise2self, krull2019noise2void, wu2020unpaired, du2020learning, zhao2021legacy, youssef2023zero} have performed denoising in an unsupervised manner without any training data or explicit noise models.

\begin{table}[t]
\caption{List of all the notations used in the paper.}
\label{notation}
\begin{center}
\begin{tabular}{p{25pt}p{200pt}}
\hline
Notation & Definition \\
\hline
$i$ & Clean and sharp image (target) \\
$o$ & Blurry and noisy observation image \\
$k$ & Blur kernel imposed on the image \\
$n$ & Additive noise imposed on the image \\
$n_s$ & Shot noise, following Poisson distribution $\mathcal{P}$ \\
$n_r$ & Read noise, following Gaussian distribution $\mathcal{G}$ \\
$n_q$ & Quantization noise, following Uniform distribution $\mathcal{U}$ \\
$K$ & Noise parameter of shot noise $n_s$ \\
$\sigma$ & Noise parameter of read noise $n_r$ \\
$qu$ & Noise parameter of quantization noise $n_q$ \\
$w$ & White balance parameter for color channels \\
$\lambda$ & Gamma correction parameter \\
$mov$ & Moving image captured from ``moving'' clip for data synthesis \\
$sta$ & Static image captured from ``static'' clip for data synthesis \\
$A$ & Exposure ratio of long- and short-exposure pixels in a Quad-Bayer data \\
\hline
$x$ & Synthetic clean dual-exposure Quad-Bayer data of QR dataset \\
$x_s$ & Synthetic clean short-exposure Quad-Bayer data of QR dataset \\
$x_l$ & Synthetic clean long-exposure Quad-Bayer data of QR dataset \\
$y$ & Synthetic clean and sharp RGB image of QR dataset \\
$y_s$ & Synthetic clean short-exposure RGB image of QR dataset, corresponding to $x_s$ \\
$y_l$ & Synthetic clean long-exposure RGB image of QR dataset, corresponding to $x_l$ \\
$q$ & Degraded Quad-Bayer data (blurry and noisy), processed from $x$ \\
\hline
$\boldsymbol{G}$ & The proposed QRNet \\
$\boldsymbol{F}$ & Fast Fourier Transform (FFT) operator \\
$\alpha$ & Frequency loss parameter \\
$\beta_1,\beta_2$ & Adam optimizer \cite{kingma2014adam} parameters \\
\hline
\end{tabular}
\end{center}
\end{table}

\noindent \textbf{Image Restoration with Burst Long- and Short-images.} Though traditional learning-based single-image deblurring and denoising methods can be potential solutions to remove artifacts, achieving practical performance is still challenging. To address this, some works proposed to use additional information, i.e., burst long- and short-exposure images. Yuan \emph{et al.} \cite{yuan2007image} first used short-exposure images to guide the estimation of blur kernels, which are then used to restore blurry images. This method was improved by the non-uniform model \cite{whyte2012non} and neural networks \cite{yamashita2017rgb}. However, their performance is restricted since the information of both inputs is not well fused. To achieve this goal, LSD2 \cite{mustaniemi2018lsd} adopted a U-Net \cite{ronneberger2015u} and used the concatenated long- and short-exposure images as the input; but it did not consider the misalignment issue between them. Chang \emph{et al.} \cite{chang2021low} used a deformable convolution \cite{dai2017deformable} since it explicitly aligned long- and short-exposure inputs. Shen \emph{et al.} extended this idea in video restoration \cite{shen2021spatial}, where optical flows were used to align both inputs. Since the motion blur scales are in a wide range for real-world photos, those methods produce visual artifacts due to the domain gap between training samples and real-world data. Also, there is a readout delay between long- and short-exposure images during capturing. To address the problems, Zhao \emph{et al.} proposed a two-stage D2HNet \cite{zhao2022d2hnet}, where the input resolution of the first stage is fixed to perform accurate deblurring. Note that, there is no readout delay problem since long- and short-exposure pixels are recorded in one image, as shown in Figure \ref{cfa} (j). The two kinds of pixels are compactly arranged so the misalignment is a controlled problem, i.e., the motion blur scales for long-exposure pixels can be estimated using neighboring short-exposure pixels. Therefore, we design a much faster approach for restoring Quad-Bayer sensor data based on these observations.

\begin{figure*}[t]
\centering
\includegraphics[width=0.95\linewidth]{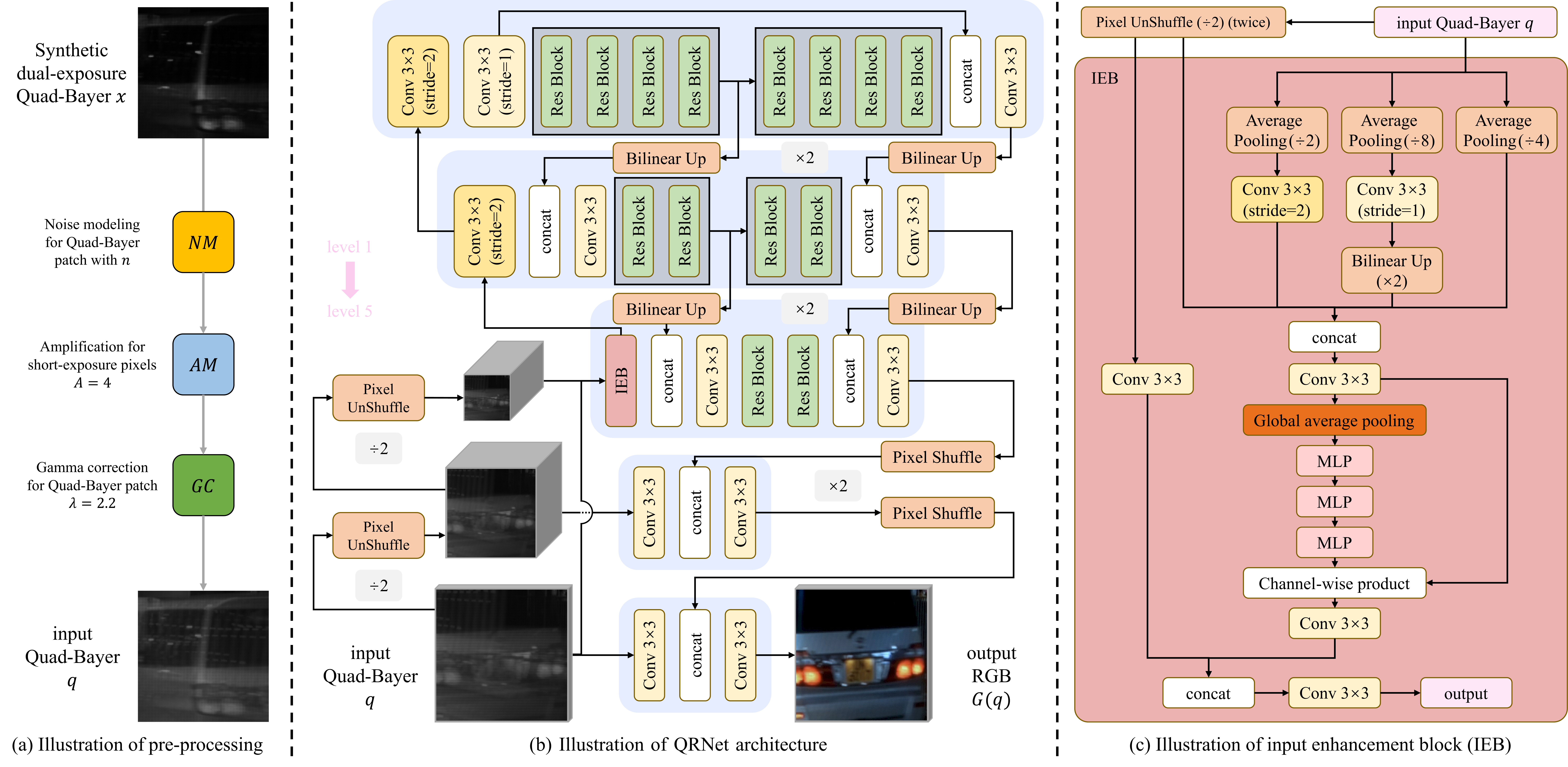}
\centering
\caption{Illustration of (a) Quad-Bayer pre-processing, (b) QRNet architecture (level 1 to 5 from top to bottom), and (c) input enhancement block.}
\label{net}
\end{figure*}

\noindent \textbf{RAW Image Restoration and RAW to RGB Mapping.} RAW is the direct output format of the camera, which contains richer information than RGB since there is no other image processing applied to it (e.g., image denoising, quantization, white balance, color correction, tone mapping, demosaicking, etc.). Thanks to the great fitting ability of neural networks, it is practical and more accurate to perform image denoising or deblurring on RAW data \cite{chen2018learning, liang2020raw}. Furthermore, researchers have developed joint image denoising and demosaicking models \cite{gharbi2016deep, zhao2019saliency, liu2020joint, xing2021end} by directly restoring a clean RGB image from a noisy RAW image. We observe that there is no explicit demosaicking method for Quad-Bayer RAW image and neural networks have the potential to map RAW to RGB. In this paper, we expand upon these applications by introducing a novel approach for jointly addressing image deblurring and denoising using a dual-exposure RAW image captured with the Quad-Bayer pattern.

\section{Methodology}

\subsection{Problem Formulation}

We aim to recover a sharp and clean RGB image from a degraded dual-exposure RAW image with the Quad-Bayer pattern (hereinafter referred to as Quad-Bayer image). We first introduce an image restoration model. Suppose that $i$ is a clean and sharp image and $o$ is a blurry and noisy observation image. Assume that $k$ is a blur kernel defined by some motion fields and $n$ is an additive noise, the formation of degraded image $o$ is:
\begin{equation}
o = k * i + n,
\label{pf1}
\end{equation}
where $*$ denotes the convolution operator. In addition, we want the model can also demosaic input Quad-Bayer data to RGB format. Assuming that $q$ is a degraded Quad-Bayer image, we aim to recover a sharp and clean RGB image $y$ by the proposed QRNet. We formulate the problem as maximizing a posteriori of the output image conditioned on input $q$ and QRNet parameters $\Theta$:
\begin{equation}
\Theta^* = \mathop{\arg\max}\limits_{\Theta} p( y | q, \Theta ),
\label{pf3}
\end{equation}
where $y$ serves as the ground truth in the optimization. All the notations are concluded in Table \ref{notation}. Details on the training process are presented below.

\subsection{Quad-Bayer Pre-processing}

There are two degradation types (blur and noise) for real dual-exposure RAW images, and we need to model them when training the QRNet. Since we collect the synthetic training Quad-Bayer data $x$ by capturing moving videos, the motion blur was modeled. To model RAW noises, we add noises to $x$ based on \cite{wei2020physics}. In addition, we propose an amplification scheme and use the gamma correction to match the intensities between Quad-Bayer input and RGB ground truth \cite{chen2018learning}, which contributes to the convergence of the QRNet. They are sequential operations as shown in Figure \ref{net} (a) and the details are concluded as:

\begin{enumerate}\setlength{\itemsep}{-0.0cm}
\item Noise modeling ($NM$): Since the noises emerge on RAW images, we add noise $n$ to the Quad-Bayer patch $x$ based on the physical model\cite{wei2020physics}, i.e., $NM (x) = x + n$;

\item Amplification ($AM$): Since the exposure times of long- and short-exposure pixels are different, we multiply short-exposure pixels with a factor $A$ to align their illuminations, i.e., $AM ( x_{short\_pixels} )= A \cdot x_{short\_pixels}$;


\item Gamma correction ($GC$): We apply a gamma correction to both long- and short-exposure pixels, which  makes the overall illumination of Quad-Bayer images comparable with target RGB images. It is implemented by $GC (x) = (x + \epsilon) ^ {(1 / \lambda)}$, where the factor $\lambda=2.2$ and $\epsilon=10^{-8}$ ($\epsilon$ circumvents the numerical instability).
\end{enumerate}

Note that the Quad-Bayer images are normalized to range of [0, 1] before performing pre-processing.

\subsection{Network Architecture}

\begin{figure}[t]
\centering
\includegraphics[width=0.95\linewidth]{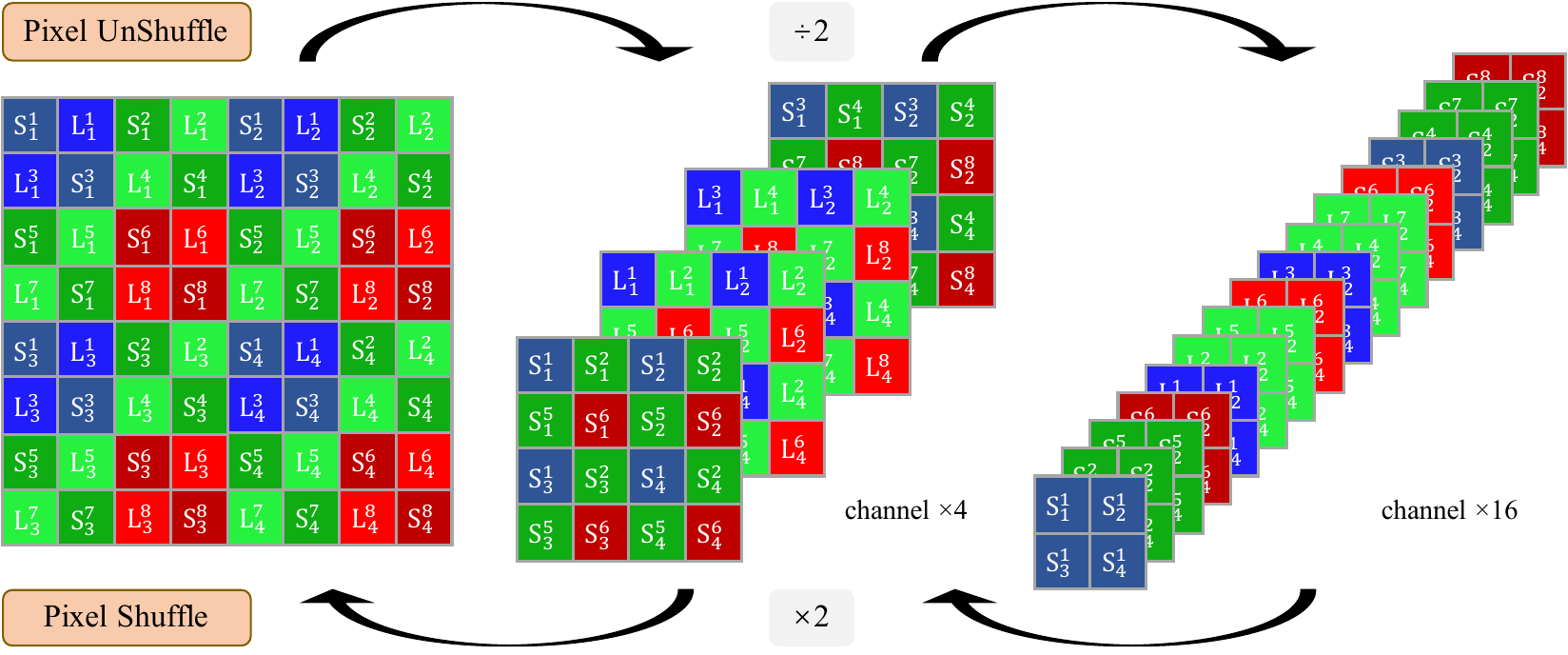}
\centering
\caption{Illustration of Pixel Unshuffle and Pixel Shuffle operations \cite{shi2016real} for Quad-Bayer images, which are invertible.}
\label{ps}
\end{figure}

The proposed Quad-Bayer to RGB mapping network (QRNet) is shown in Figure \ref{net} (b). It is an end-to-end network that recovers an RGB image from an input Quad-Bayer image. It is also a multi-level architecture, and we define levels 1 to 5 from top to bottom based on the feature or image resolution, e.g., level 5 has the original resolution. We describe the designs in detail as follows:

\noindent \textbf{Data Re-arrangement.} Taking into account the Quad-Bayer layout, the short- and long-exposure pixels are densely packed within a single channel. Directly employing convolutional layers on this layout poses challenges for feature extraction, as the layers must additionally discern the spatial relationships between pixels captured at two different exposure levels. To mitigate this issue, we perform the Pixel Unshuffle operation \cite{shi2016real} twice on the input Quad-Bayer data. This operation segregates pixels based on their exposure times and colors into distinct channels, as illustrated in Figure \ref{ps}.

\noindent \textbf{Input enhancement block.} We notice that neighboring pixels of Pixel Unshuffle downsampled features are not continuous (e.g., pixels $S^1_1$ and $S^1_2$ in Figure \ref{ps}). This sampling method might cause jagged artifacts. To alleviate it, we propose an input enhancement block (IEB), as shown in Figure \ref{net} (c). In addition to Pixel Unshuffle operations, we include three more average pooling branches. It merges the intensity of neighboring pixels of original Quad-Bayer data that has better continuity. Then, we merge the three average pooling downsampled data and Pixel Unshuffle downsampled data by channel attention, inspired by \cite{hu2018squeeze}. After that, we combine the channel attention features with features extracted from Pixel Unshuffle downsampled data by another convolutional layer.

\noindent \textbf{Multi-level feature extraction.} Based on the result from IEB, we further define a three-level structure (upper three levels of Figure \ref{net} (b)) to fuse the information from both short- and long-exposure pixels. We use residual blocks \cite{he2016deep} at levels 1 to 3 to improve the feature representation. For levels 2 and 3, the information from the previous level is combined before entering the residual blocks. All the convolutional layers adopt 3$\times$3 kernels. Since the majority of operations are performed at levels 1 to 3, which have small image or feature resolutions, the architecture is efficient at inference.

\noindent \textbf{Feature interactions between levels.} Levels 1 to 5 of QRNet are related by upsampling the features from the previous level. For levels 1 to 3, there are two connections between every two neighboring levels at the start and the tail, respectively. The connection at the start combines the information from the previous level and the input of the next level. Then, we use a convolutional layer to emphasize the feature fusion and suppress the number of channels. It makes the current level possess multi-scale features, which enlarges the perceptive field and enhances the feature representation. The tail connection passes the whole information from the previous level to the next level. For levels 4 and 5, the relation is implemented by Pixel Shuffle upsampling and concatenation, which corresponds to the downsampling operations of the input pyramid. It also improves the image restoration results since every level can exploit multi-scale features.

\subsection{Loss Functions}

Likewise existing image restoration methods, we exploit the L1 loss as reconstruction loss to make the generated image close to ground truth. It is defined as:
\begin{equation}
L_{recon} = \mathbb{E}[||\boldsymbol{G}(q) - y||_1],
\label{loss1}
\end{equation}
where $\boldsymbol{G}(q)$ and $y$ represent the QRNet output and the ground truth RGB image, respectively.

To further reconstruct high-frequency parts of an image, we use a frequency loss \cite{cho2021rethinking} at the training. It measures the L1 distances between the generated image and the ground truth in the frequency domain as follows:
\begin{equation}
L_{freq} = \mathbb{E}[||\boldsymbol{F} ( \boldsymbol{G}(q) ) - \boldsymbol{F} ( y )||_1],
\label{loss2}
\end{equation}
where $\boldsymbol{F}$ is the fast Fourier transform (FFT) operator that transforms the image from the spatial domain into the frequency domain. Combining the two loss functions, we define the overall loss as:
\begin{equation}
L_{overall} = L_{recon} + \alpha L_{freq},
\label{loss3}
\end{equation}
where $\alpha$ is a trade-off parameter.

\begin{figure*}[t]
\centering
\includegraphics[width=0.95\linewidth]{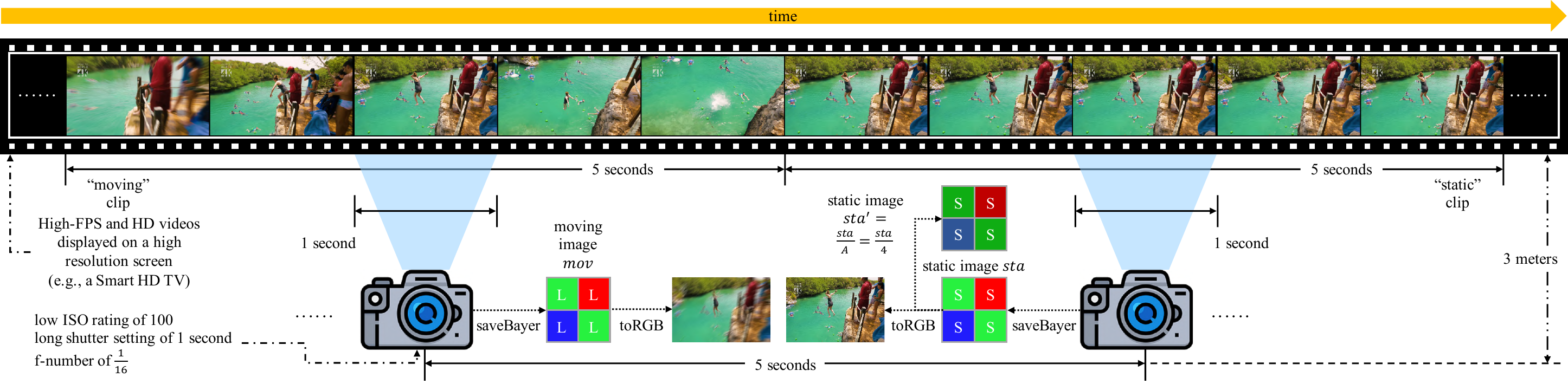}
\centering
\caption{Illustration of the data capturing workflow for a moving image and a static image from an edited video.}
\label{dataset1}
\end{figure*}

\begin{figure*}[t]
\centering
\includegraphics[width=0.95\linewidth]{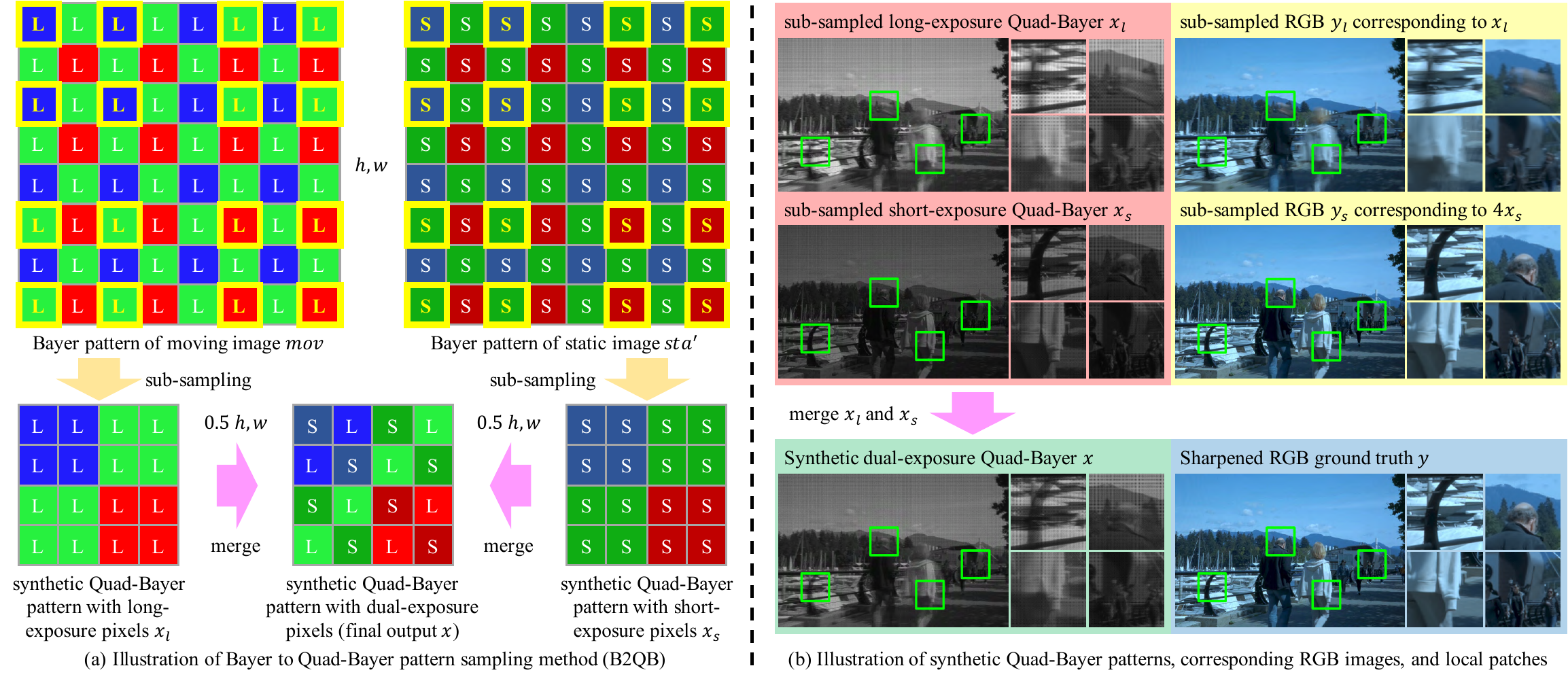}
\centering
\caption{Illustration of (a) Bayer to Quad-Bayer pattern sampling method (B2QB); (b) synthetic Quad-Bayer patterns, their corresponding RGB images, and four local patches extracted from the full resolution images. To better illustrate those images, we post-process them to have the same illuminations.}
\label{dataset2}
\end{figure*}

\section{Dataset Collection}

In this section, we introduce the data collection details of the Quad-Bayer to RGB mapping dataset (QR dataset). The motivation that we synthesize training pairs instead of collecting them by Quad-Bayer module is in three folds:

\begin{enumerate}\setlength{\itemsep}{-0.0cm}
\item It is tedious and time-consuming to change the settings for Quad-Bayer modules;

\item There are no effective demosaicking methods for Quad-Bayer sensor data (i.e., hard to obtain corresponding RGB images as the target for the proposed algorithm);

\item A perfect alignment is hard to obtain when capturing multiple Quad-Bayer and RGB pairs.

\end{enumerate}


Therefore, we propose a data collection pipeline to create aligned and paired Quad-Bayer and RGB images. It includes four sequential steps:

\noindent \textbf{Preparation.} Considering the motivation, to build a sufficient dataset to learn the Quad-Bayer to RGB mapping, we capture Bayer RAW images first and then sample pixels to obtain their Quad-Bayer counterparts (i.e., B2QB scheme). Since Quad-Bayer RAW data includes two exposures, we need the paired long- and short-exposure Bayer RAW images. To capture these images, we first download 30 HD videos (2160p) from YouTube, which cover different real-world scenes (e.g., country, city, mountain, forest, etc.) and are almost noise-free.


\noindent \textbf{Data modeling.} To better capture images, we post-process these videos to paired ``moving'' and ``static'' clips, for capturing long- and short-exposure images, respectively. In detail, we cut every original video into clips of 2.5 seconds with a sampling step of 45 seconds. For each clip, we interpolate it 2 times longer, i.e., 5 seconds. It is called a ``moving'' clip. From the ``moving'' clip, the middle frame is extracted and repeated to form a ``static'' clip of 5 seconds and the same FPS as the corresponding ``moving'' clip. These paired ``moving'' and ``static'' clips form a group and are composed into one single video for capturing, as shown in Figure \ref{dataset1}.



\noindent \textbf{Data capturing.} To control the environment, we capture the edited clips on a high-resolution monitor with a DSLR camera, Nikon D5200, which captures RAW images with 14 bits. All other lighting sources are closed except the monitor. The display fully occupies the camera’s FOV. We use a low ISO rating of 100, a shutter setting of 1 second, and an f-number of $\frac{1}{16}$ to get rid of noises as many as possible. A long shutter speed will induce overexposure and extreme blur and a short shutter speed will result in gloomy and noisy images, while a 1-second shutter speed is a trade-off setting. The RAW images taken under such conditions are almost noise-free. The camera is placed 3 meters away from the monitor with a tripod and takes an image every 5 seconds controlled by a programmable shutter release. The workflow of capturing a pair of moving image $mov$ and static image $sta$ is illustrated in Figure \ref{dataset1}. Since the overall exposure times and ISO ratings for $mov$ and $sta$ are the same, their brightness of them is the same. However, long- and short-exposure pixels of real Quad-Bayer images should have different brightness due to different exposure times. Therefore, we apply an exposure ratio $A$ to the static image $sta$, i.e., the processed ${sta}^{\prime}=\frac{sta}{A}$. In this paper, we set $A=4$ according to the real Quad-Bayer device.

\begin{table*}[t]
\begin{center}
\caption{The comparisons of QRNet and other methods based on dual-exposure Quad-Bayer images. Different rows include the results from different methods and the last row is the proposed QRNet. Columns 1-4 include the methods, input data structure, MACs (on 512$\times$512 resolution), and running time (second per 512$\times$512 patch), respectively. Columns 5-7 PSNR \& SSIM results of different noise parameters. The \textcolor{red}{red}, \textcolor{blue}{blue}, and \textcolor{green}{green} colors denote the best, the second highest and the third highest results, respectively.}
\label{table_sota1}
\begin{tabular}{l|c|c|c|cc|cc|cc}
\hline
\multirow{2}{*}{Method} & \multirow{2}{*}{Input data structure} & MACs & Running time & \multicolumn{2}{c|}{$K=0.25$, $\sigma=5$} & \multicolumn{2}{c|}{$K=0.5$, $\sigma=5$} & \multicolumn{2}{c}{$K=0.75$, $\sigma=5$} \cr & & (512$\times$512) & (sec / 512$\times$512) & PSNR & SSIM & PSNR & SSIM & PSNR & SSIM \\
\hline
LSD2 \cite{mustaniemi2018lsd} & dual-exposure Quad-Bayer & \textcolor{green}{113.433G} & 0.0079 & 32.21 & 0.9264 & 32.67 & 0.9317 & 32.81 & 0.9333 \\
MWCNN \cite{liu2018multi} & dual-exposure Quad-Bayer & 301.671G & 0.1514 & 31.85 & 0.9193 & 32.28 & 0.9244 & 32.47 & 0.9278 \\
SGN \cite{gu2019self} & dual-exposure Quad-Bayer & \textcolor{blue}{51.258G} & 0.0113 & 32.00 & 0.9227 & 32.46 & 0.9281 & 32.64 & 0.9299 \\
MIRNet \cite{zamir2020learning} & dual-exposure Quad-Bayer & 2.607T & 0.6869 & 32.18 & 0.9275 & 32.72 & 0.9326 & 32.87 & 0.9339 \\
MPRNet \cite{zamir2021multi} & dual-exposure Quad-Bayer & 2.642T & 0.1435 & 32.24 & 0.9267 & 32.41 & 0.9301 & 32.81 & 0.9323 \\
SRN \cite{tao2018scale} & dual-exposure Quad-Bayer & 343.004G & 0.0545 & 32.42 & 0.9296 & 32.86 & 0.9345 & 33.01 & 0.9358 \\
DeblurGAN \cite{kupyn2018deblurgan} & dual-exposure Quad-Bayer & 205.445G & 0.0111 & 32.55 & 0.9303 & 32.93 & 0.9336 & 33.15 & 0.9360 \\
DMPHN \cite{zhang2019deep} & dual-exposure Quad-Bayer & 202.458G & 0.0995 & 32.51 & 0.9303 & 32.93 & 0.9344 & 33.10 & 0.9359 \\
MIMOUNet \cite{cho2021rethinking} & dual-exposure Quad-Bayer & 300.205G & 0.0394 & \textcolor{green}{32.71} & \textcolor{green}{0.9332} & \textcolor{green}{33.09} & \textcolor{green}{0.9366} & \textcolor{green}{33.31} & \textcolor{green}{0.9387} \\
MIMOUNet++ \cite{cho2021rethinking} & dual-exposure Quad-Bayer & 687.105G & 0.0831 & \textcolor{blue}{32.82} & \textcolor{blue}{0.9350} & \textcolor{blue}{33.26} & \textcolor{blue}{0.9386} & \textcolor{blue}{33.40} & \textcolor{blue}{0.9398} \\
\hline
QRNet & dual-exposure Quad-Bayer & \textcolor{red}{34.628G} & 0.0178 & \textcolor{red}{33.24} & \textcolor{red}{0.9403} & \textcolor{red}{33.61} & \textcolor{red}{0.9428} & \textcolor{red}{33.81} & \textcolor{red}{0.9445} \\
\hline
\end{tabular}
\end{center}
\end{table*}

\noindent \textbf{B2QB Scheme.} After capturing, we obtain pairs of blurry and sharp images with trivial noises and different brightness, i.e., $mov$ and $sta^{\prime}$. They are regarded as long- and short-exposure images, respectively. Then, we synthesize the corresponding \emph{Quad-Bayer data} following the B2QB scheme as illustrated in Figure \ref{dataset2} (a), which yields downsized images with half resolution of DSLR outputs, denoted as $x_l$ and $x_s$. Finally, we combine the long- and short-exposure Quad-Bayer data to obtain the dual-exposure Quad-Bayer data $x$. For \emph{target RGB images}, we propose to demosaic the Bayer patterns $sta$ and apply the same sampling scheme in order to minimize the edge differences. In detail, we first apply white balance. The gain for green channel $w_g$ is fixed to 1. The gains for red channel $w_r$ and blue channel $w_b$ are sampled uniformly from [1.6, 1.8] and [1.4, 1.6], respectively. Then, we apply gamma correction with $\lambda=2.2$, which is an exponential calculation. Next, we demosaic them with \cite{menon2006demosaicing}. After that, we apply the MATLAB ``imsharpen'' tool to post-process them. Finally, we apply the same sampling scheme and obtain $y$. We also illustrate the transformed RGB images with the same demosaicking and sampling schemes corresponding to $x_l$ and $4x_s$ (i.e., the intensity of $4x_s$ is four times larger than $x_s$) in Figure \ref{dataset2} (b), respectively.

The overall synthesis workflow results in 731 tuples ($x_s,x_l,x,y_s,y_l,y$), where every dual-exposure Quad-Bayer image ($x$) and sharpened RGB image ($y$) forms a pair. They have a resolution of 3016$\times$2008. They are divided into 701 training tuples and 30 validation tuples.

In addition, we capture 24 real Quad-Bayer images with the Sony IMX 586 Quad-Bayer module for testing. The data arrangement is identical to our synthetic data. The captured images are all indoor scenes. The image resolution is 4608$\times$3456 and the data is in 10 bits.

\section{Experiment}

\subsection{Implementation Details}

We train the QRNet with 200 epochs and each epoch has 2500 iterations. The batch size is 16. To facilitate the training, we select four full-resolution images first and then crop four 320$\times$320 patches from each full-resolution image. The learning rate is initialized as $1\times10^{-4}$. During the 100-th epoch to the 200-th epoch, the learning rate is linearly decreased to $5\times10^{-7}$. The frequency loss parameter $\alpha$ is set to 0.01. We use Adam optimizer \cite{kingma2014adam} with $\beta_1=0.5$, $\beta_2=0.999$. All trainable parameters of QRNet are initialized to random values from the standard normal distribution. We implement the QRNet with PyTorch 1.0.0 and CUDA 9.0. We train it on 4 NVIDIA TITAN Xp GPUs. It takes approximately 30 hours to complete the whole training process.

\begin{table*}[t]
\begin{center}
\caption{The comparisons of QRNet and MIMOUNet++ \cite{cho2021rethinking} based on dual-exposure Quad-Bayer images. More noise parameter settings are included. The \textcolor{red}{red} color highlights the better results among QRNet and MIMOUNet++.}
\label{table_sota2}
\begin{tabular}{l|cc|cc|cc|cc|cc|cc}
\hline
\multirow{2}{*}{Method} & \multicolumn{2}{c|}{$K=0.75$, $\sigma=5$} & \multicolumn{2}{c|}{$K=5$, $\sigma=5$} & \multicolumn{2}{c|}{$K=0.75$, $\sigma=3$} & \multicolumn{2}{c|}{$K=5$, $\sigma=3$} & \multicolumn{2}{c|}{$K=0.75$, $\sigma=1$} & \multicolumn{2}{c}{$K=5$, $\sigma=1$} \cr & PSNR & SSIM & PSNR & SSIM & PSNR & SSIM & PSNR & SSIM & PSNR & SSIM & PSNR & SSIM \\
\hline
MIMOUNet++ \cite{cho2021rethinking} & 33.40 & 0.9398 & 33.84 & 0.9439 & 34.53 & 0.9486 & 35.29 & 0.9536 & 35.85 & 0.9578 & 37.38 & 0.9653 \\
\hline
QRNet & \textcolor{red}{33.81} & \textcolor{red}{0.9445} & \textcolor{red}{34.19} & \textcolor{red}{0.9471} & \textcolor{red}{34.91} & \textcolor{red}{0.9523} & \textcolor{red}{35.60} & \textcolor{red}{0.9563} & \textcolor{red}{36.27} & \textcolor{red}{0.9617} & \textcolor{red}{37.65} & \textcolor{red}{0.9672} \\
\hline
\end{tabular}
\end{center}
\end{table*}

\begin{figure*}[t]
\centering
\includegraphics[width=0.95\linewidth]{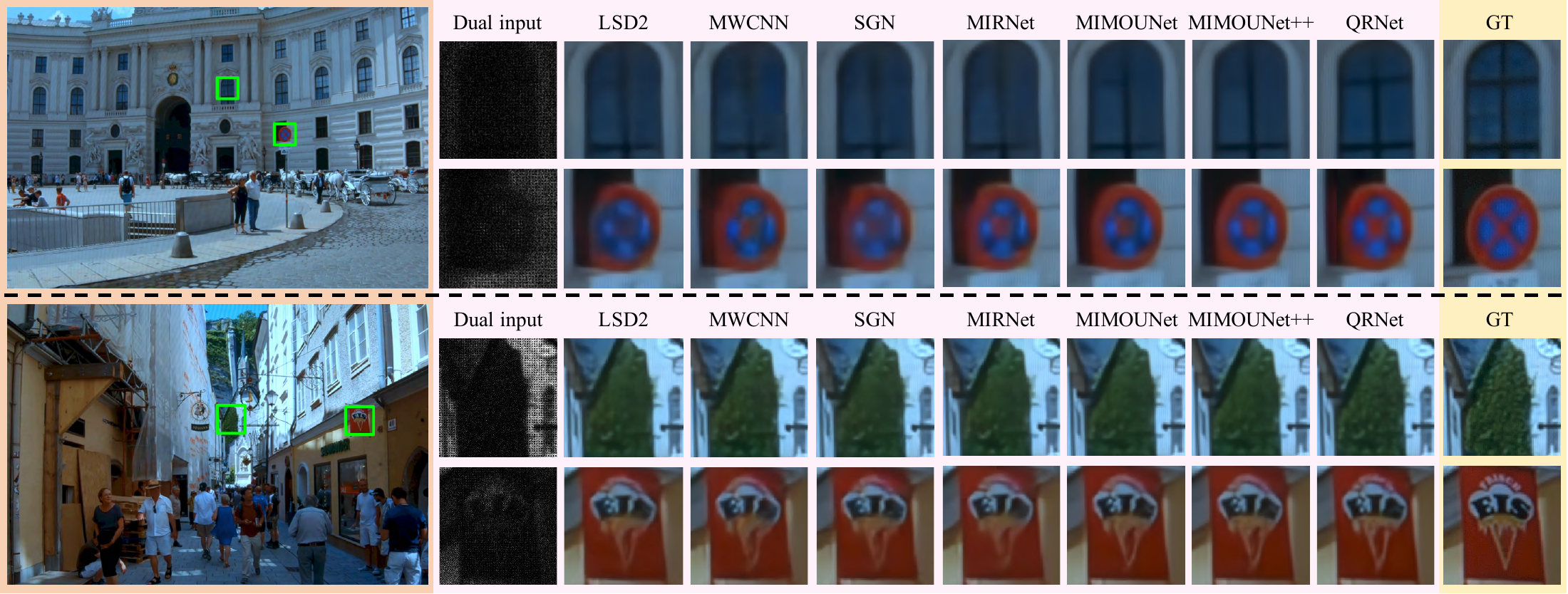}
\centering
\caption{Illustration of restoration results of QRNet and other methods based on dual-exposure Quad-Bayer images. The noise parameter is $K=0.25, \sigma=5$. The left is full-resolution ground truth images. On the right, input dual-exposure Quad-Bayer image patches and generated image patches are in the pink background. Ground truth image patches are in the yellow background.}
\label{fig_sota1}
\end{figure*}

\begin{figure*}[t]
\centering
\includegraphics[width=0.95\linewidth]{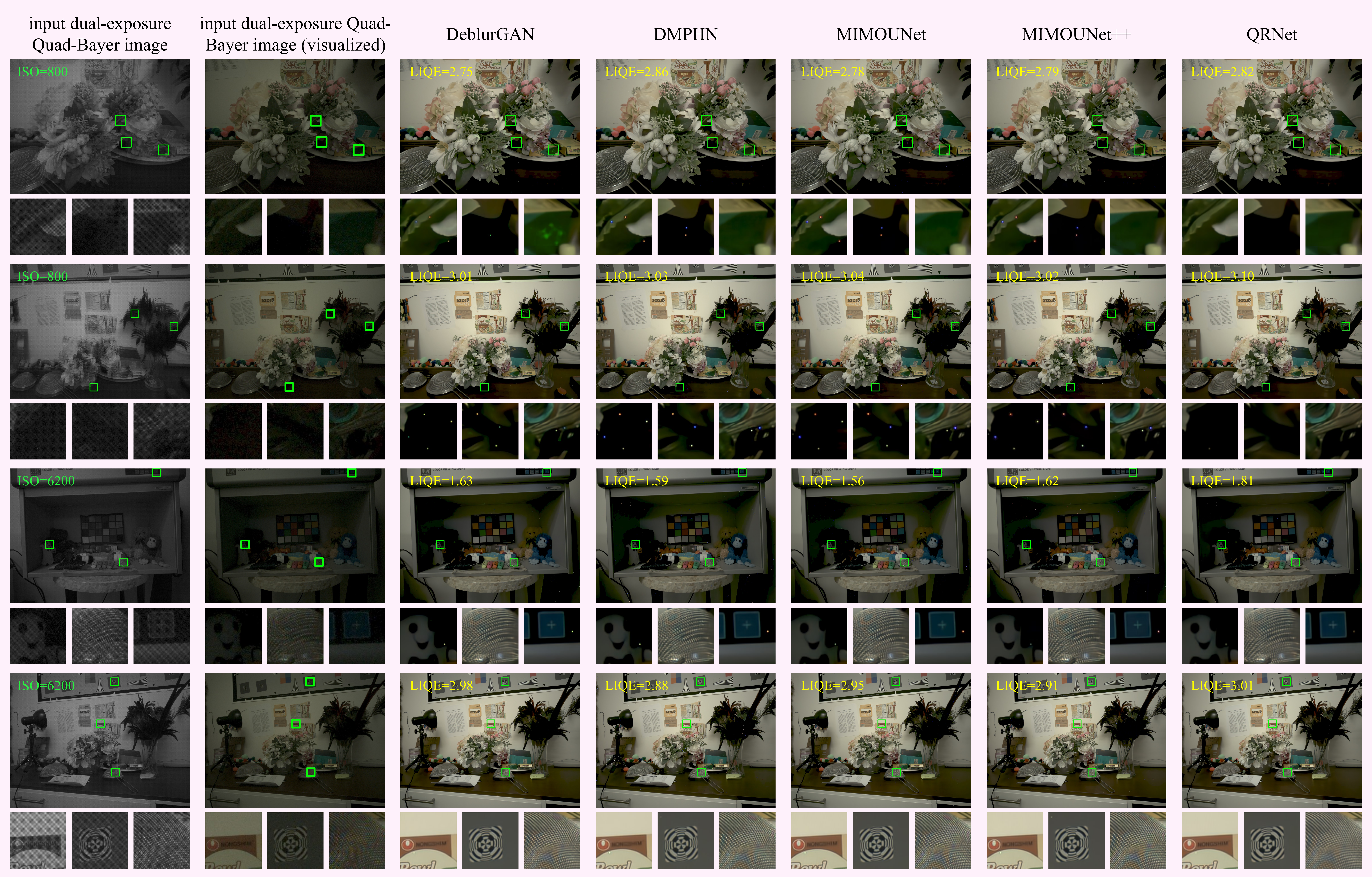}
\centering
\caption{Illustration of 4 real input Quad-Bayer images and restoration results from different methods. For every full-resolution image, three patches are extracted and named $\sharp$1, $\sharp$2, and $\sharp$3. The left column includes input Quad-Bayer images, where ISO values are shown at the top left corner. The second-left column includes the visualization results by demosaicking the Bayer counterparts of input images. The other columns include restoration results from DeblurGAN \cite{kupyn2018deblurgan}, DPMHN \cite{zhang2019deep}, MIMOUNet \cite{cho2021rethinking}, MIOMUNet++ \cite{cho2021rethinking}, and QRNet, where LIQE \cite{zhang2023liqe} values are shown at the top left corner.}
\label{fig_sota2}
\end{figure*}

\subsection{Experiment Details}

\noindent \textbf{Experiment settings.} To perform a comprehensive and fair comparison, we first compare the proposed QRNet with other methods from degraded dual-exposure Quad-Bayer to clean and sharp RGB $y$. Then, we compare them on four more pipelines:

\begin{enumerate}\setlength{\itemsep}{-0.0cm}
\item Short-exposure-RAW-based methods: From degraded short-exposure Quad-Bayer to clean and sharp RGB $y$;

\item Long-exposure-RAW-based methods: From degraded long-exposure Quad-Bayer to clean and sharp RGB $y$;

\item Short-exposure-RGB-based methods: From degraded short-exposure RGB to clean and sharp RGB $y$;

\item Long-exposure-RGB-based methods: From degraded long-exposure RGB to clean and sharp RGB $y$.
\end{enumerate}

For the training of other methods, the data selection scheme, training epochs, and batch size are identical to QRNet for fair comparisons. Since the synthetic tuples $x_s,x_l,x,y_s,y_l,y$ are almost noise-free, we add noises with different levels to the synthetic data for benchmarking, which matches the real-world application scenarios.

\noindent \textbf{Noise modeling.} We add the noise $n$ in the RAW space based on the physical model \cite{wei2020physics}, which can be represented as:
\begin{equation}
\begin{aligned}
& n = n_s + n_r + n_q, \\
& x_* + n_s \sim \mathcal{P} (x_* \cdot K) / K, \\
& n_r \sim \mathcal{G} (0, \sigma^2), \\
& n_q \sim \mathcal{U} (-1 / 2qu, 1 / 2qu),
\label{experiment_setting1}
\end{aligned}
\end{equation}
where $x_*$ is a clean RAW image. $n_s$, $n_r$, and $n_q$ are shot, read, and quantization noises, respectively. $\mathcal{P}$, $\mathcal{G}$, and $\mathcal{U}$ denote Poisson, Gaussian, and Uniform distributions, respectively. Here $K$ and $\sigma$ are noise model hyper-parameters, where we set $K=0.25$ or $0.5$ or $0.75$ and $\sigma=5$ in the experiment. They simulate different noise levels. $qu$ is a static quantization step. After adding noises to the RAW image, we multiply the exposure ratio $A$ with short-exposure pixels.

Note if inputs are RGB images ($y_{s}$ or $y_{l}$), we first convert them to RAW space to add noises and then transform them back to RGB images \cite{brooks2019unprocessing} at training.
\begin{equation}
\begin{aligned}
& \widehat{y_s} = RAW2RGB (A \cdot (RGB2RAW (y_s) / A + n)), \\
& \widehat{y_l} = RAW2RGB (RGB2RAW (y_l) + n),
\label{experiment_setting2}
\end{aligned}
\end{equation}
where $\widehat{y_s}$ and $\widehat{y_l}$ are noisy versions of $y_s$ and $y_l$ for training. $RAW2RGB$ contains demosaicking, white balance, and gamma correction and $RGB2RAW$ contains the inverse transformations. $n$ is the added noises as shown in Equation \ref{experiment_setting1}.

\noindent \textbf{Evaluation metrics.} We employ PSNR and SSIM metrics \cite{wang2004image} to assess the pairwise image quality across 30 validation tuples within the QR dataset. Additionally, we determine the Multiply-Accumulate (MACs) using the PyTorch-OpCounter \cite{opcounter} library. The processing time for a 512$\times$512 patch on an NVIDIA RTX 3090 GPU is calculated by averaging the time taken to forward 10000 patches for ten repetitions. These computations were performed using PyTorch 1.7.0 and CUDA 11.0.

\subsection{Evaluation with Dual-exposure-based Methods} \label{subsection_dual_exposure}

In this section, we evaluate the image restoration quality of QRNet and state-of-the-art methods including LSD2 \cite{mustaniemi2018lsd}, MWCNN \cite{liu2018multi}, SGN \cite{gu2019self}, MIRNet \cite{zamir2020learning}, SRN \cite{tao2018scale}, DeblurGAN \cite{kupyn2018deblurgan}, DMPHN \cite{zhang2019deep}, MIMOUNet, and MIMOUNet++ \cite{cho2021rethinking} based on dual-exposure Quad-Bayer data.

First, we conduct the quantitative analysis on the QR validation set, where the results are concluded in Table \ref{table_sota1}. It is clear that QRNet has the least MACs, and the highest PSNR and SSIM across all the methods. QRNet also outperforms all other methods under different noise levels, which demonstrates its stability. Compared with the state-of-the-art method MIMOUNet++, QRNet only has 5\% MACs of it and obtains 0.32 higher PSNR under $K=0.25, \sigma=5$ setting. Compared with the other fast method SGN, QRNet obtains 1.14 higher PSNR and 0.0158 higher SSIM under $K=0.25, \sigma=5$ setting. To further demonstrate the superior image restoration quality of QRNet, we conduct 5 more experiments as concluded in Table \ref{table_sota2}. QRNet obtains better PSNR and SSIM values than the state-of-the-art method MIMOUNet++ even under more different noise levels. The experiments demonstrate the QRNet architecture is advanced for restoring Quad-Bayer sensor data.

We show some generated images in Figure \ref{fig_sota1}. The results from QRNet are sharper, e.g., the crossbar of the window is recovered by QRNet while other methods do not in the first row; there are fewer color bleeding artifacts for QRNet in the second row; the textures and edges are more obvious for QRNet in the third and fourth rows. They show the QRNet has a better ability to restore details and textures than existing methods under the same noise parameter in the training.

Secondly, we evaluate the QRNet and other methods \cite{kupyn2018deblurgan, zhang2019deep, cho2021rethinking} on real Quad-Bayer images captured by the IMX586 module. The noise parameter is $K=5, \sigma=1$ for all the methods at training, which is estimated on real Quad-Bayer images based on \cite{wei2020physics}. We show two blurry samples (top two rows) and two highly noisy samples (bottom two rows) in Figure \ref{fig_sota2}. On one hand, since there is no ground truth for real Quad-Bayer images, we adopt the non-reference image quality evaluator LIQE \cite{zhang2023liqe} to measure the quality of all generated images, which better aligns with mean opinion scores (MOSs) than previous evaluators. The LIQE values are illustrated on the top left corner of every generated image in Figure \ref{fig_sota2}. For all 4 samples, QRNet has higher LIQE values than other methods, which denotes that QRNet's results better match perceptual quality. On the other hand, we notice that other methods produce more abnormal pixels than QRNet, e.g., patches $\sharp$1-$\sharp$2 of row 1, patches $\sharp$1-$\sharp$3 of row 2, and patch $\sharp$3 of row 3. Compared with other methods, QRNet produces less color artifacts, e.g., patch $\sharp$2 of row 3 and patch $\sharp$3 of row 4. In addition, QRNet generates clearer words, e.g., ``NONGSHIM'' in patch $\sharp$1 of row 4.

From the experiment results, we observe that QRNet outperforms existing methods on both synthetic and real dual-exposure Quad-Bayer images. We assume the QRNet architecture (including the input enhancement block, multi-level feature extraction, and inter-level feature interactions) can better restore RAW images with the Quad-Bayer pattern compared with existing methods.

\begin{table}[t]
\begin{center}
\caption{The comparisons of QRNet and other methods on dual-exposure Quad-Bayer RAW image restoration.}
\label{table_raw}
\resizebox{88mm}{19mm}{
\begin{tabular}{l|cc|cc|cc}
\hline
\multirow{2}{*}{Method} & \multicolumn{2}{c|}{$K=0.25$, $\sigma=5$} & \multicolumn{2}{c|}{$K=0.5$, $\sigma=5$} & \multicolumn{2}{c}{$K=0.75$, $\sigma=5$} \cr & PSNR & SSIM & PSNR & SSIM & PSNR & SSIM \\
\hline
LSD2 \cite{mustaniemi2018lsd} & 39.20 & 0.9907 & 40.26 & 0.9918 & 40.75 & 0.9923 \\
MWCNN \cite{liu2018multi} & 39.14 & 0.9907 & 40.23 & 0.9919 & 40.73 & 0.9923 \\
SGN \cite{gu2019self} & 38.72 & 0.9899 & 40.01 & 0.9915 & 40.49 & 0.9919 \\
MIRNet \cite{zamir2020learning} & 39.03 & 0.9905 & 40.15 & 0.9917 & 40.64 & 0.9922 \\
MPRNet \cite{zamir2021multi} & 39.09 & 0.9906 & 40.42 & 0.9922 & 40.92 & \textcolor{green}{0.9926} \\
SRN \cite{tao2018scale} & 39.63 & 0.9916 & 40.47 & \textcolor{green}{0.9923} & 41.19 & 0.9930 \\
DeblurGAN \cite{kupyn2018deblurgan} & 39.07 & 0.9906 & 40.05 & 0.9917 & 40.58 & 0.9922 \\
DMPHN \cite{zhang2019deep} & 39.17 & 0.9908 & 39.98 & 0.9916 & 40.55 & 0.9921 \\
MIMOUNet \cite{cho2021rethinking} & \textcolor{blue}{39.64} & \textcolor{blue}{0.9916} & \textcolor{green}{40.70} & \textcolor{blue}{0.9926} & \textcolor{green}{41.22} & \textcolor{blue}{0.9931} \\
MIMOUNet++ \cite{cho2021rethinking} & \textcolor{green}{39.62} & \textcolor{green}{0.9915} & \textcolor{blue}{40.72} & \textcolor{blue}{0.9926} & \textcolor{blue}{41.27} & \textcolor{blue}{0.9931} \\
\hline
QRNet & \textcolor{red}{39.88} & \textcolor{red}{0.9919} & \textcolor{red}{40.88} & \textcolor{red}{0.9928} & \textcolor{red}{41.35} & \textcolor{red}{0.9932} \\
\hline
\end{tabular}
}
\end{center}
\end{table}

\subsection{Evaluation on Quad-Bayer RAW Image Restoration}

In this section, we conduct experiments on joint deblurring and denoising in the Quad-Bayer RAW domain. The input is a blurry and noisy Quad-Bayer RAW image while the output is a sharp and clean one. Compared with RAW to RGB mapping, restoring RAW images enables the adjustment of the following ISP steps such as white balance, tone mapping, color adjustment, etc. Specifically, we compare QRNet with methods described in Section \ref{subsection_dual_exposure} while the target changes to clean and sharp Quad-Bayer data. The quantitative results are concluded in Table \ref{table_raw}, where QRNet still achieves higher PSNR and SSIM values than other methods under 3 different noise levels. It further demonstrates that QRNet architecture can better restore Quad-Bayer images even having less computation costs.

\begin{table*}[t]
\begin{center}
\caption{The comparisons of QRNet and other baselines based on short-exposure images.}
\label{table_com1}
\begin{tabular}{l|c|c|c|cc|cc|cc}
\hline
\multirow{2}{*}{Method} & \multirow{2}{*}{Input data structure} & MACs & Running time & \multicolumn{2}{c|}{$K=0.25$, $\sigma=5$} & \multicolumn{2}{c|}{$K=0.5$, $\sigma=5$} & \multicolumn{2}{c}{$K=0.75$, $\sigma=5$} \cr & & (512$\times$512) & (sec / 512$\times$512) & PSNR & SSIM & PSNR & SSIM & PSNR & SSIM \\
\hline
IRCNN \cite{zhang2017learning} & short-exposure Quad-Bayer & \textcolor{green}{54.459G} & 0.0026 & 28.99 & 0.8914 & 29.23 & 0.8930 & 29.42 & 0.8955 \\
DnCNN \cite{zhang2017beyond} & short-exposure Quad-Bayer & 162.001G & 0.0057 & 28.82 & 0.8917 & 29.26 & 0.8939 & 29.12 & 0.8936 \\
MemNet \cite{tai2017memnet} & short-exposure Quad-Bayer & 194.986G & 0.0305 & 15.46 & 0.7517 & 15.26 & 0.7490 & 15.62 & 0.7588 \\
MWCNN \cite{liu2018multi} & short-exposure Quad-Bayer & 301.671G & 0.1514 & 31.71 & 0.9177 & 31.89 & 0.9187 & 31.98 & 0.9197 \\
SGN \cite{gu2019self} & short-exposure Quad-Bayer & \textcolor{blue}{51.258G} & 0.0113 & \textcolor{green}{31.78} & 0.9183 & 32.16 & 0.9231 & 32.27 & 0.9240 \\
DSWN \cite{liu2020densely} & short-exposure Quad-Bayer & 102.283G & 0.0116 & 31.24 & 0.9114 & 31.48 & 0.9141 & 31.57 & 0.9134 \\
ADNet \cite{tian2020attention} & short-exposure Quad-Bayer & 137.158G & 0.0066 & 16.71 & 0.7870 & 16.98 & 0.7905 & 17.10 & 0.7954 \\
RIDNet \cite{anwar2019real} & short-exposure Quad-Bayer & 439.379G & 0.0182 & 30.09 & 0.9062 & 30.34 & 0.9088 & 30.29 & 0.9084 \\
MIRNet \cite{zamir2020learning} & short-exposure Quad-Bayer & 2.607T & 0.6869 & 31.36 & 0.9079 & \textcolor{green}{32.17} & \textcolor{green}{0.9237} & \textcolor{green}{32.34} & \textcolor{green}{0.9261} \\
MPRNet \cite{zamir2021multi} & short-exposure Quad-Bayer & 2.642T & 0.1435 & 30.19 & 0.9089 & 30.13 & 0.9060 & 29.65 & 0.8969 \\
HINet \cite{chen2021hinet} & short-exposure Quad-Bayer & 769.487G & 0.0315 & 24.15 & 0.8315 & 27.62 & 0.8810 & 21.92 & 0.8054 \\
QRNet & short-exposure Quad-Bayer & \textcolor{red}{34.628G} & 0.0178 & \textcolor{blue}{32.89} & \textcolor{blue}{0.9372} & \textcolor{blue}{33.19} & \textcolor{blue}{0.9400} & \textcolor{blue}{33.33} & \textcolor{blue}{0.9411} \\
\hline
IRCNN \cite{zhang2017learning} & short-exposure RGB & \textcolor{green}{54.794G} & 0.0026 & 29.86 & 0.8927 & 30.03 & 0.8942 & 30.16 & 0.8945 \\
DnCNN \cite{zhang2017beyond} & short-exposure RGB & 162.336G & 0.0059 & 29.89 & 0.8939 & 30.04 & 0.8946 & 30.21 & 0.8955 \\
MemNet \cite{tai2017memnet} & short-exposure RGB & 195.140G & 0.0309 & 16.59 & 0.7723 & 16.86 & 0.7829 & 16.73 & 0.7850 \\
MWCNN \cite{liu2018multi} & short-exposure RGB & 301.671G & 0.1517 & 31.20 & 0.9134 & 31.45 & 0.9160 & 31.54 & 0.9172 \\
SGN \cite{gu2019self} & short-exposure RGB & \textcolor{blue}{53.774G} & 0.0116 & 31.48 & 0.9157 & 31.81 & 0.9205 & 31.96 & 0.9221 \\
DSWN \cite{liu2020densely} & short-exposure RGB & 104.800G & 0.0113 & 30.71 & 0.9063 & 31.19 & 0.9105 & 31.21 & 0.9116 \\
ADNet \cite{tian2020attention} & short-exposure RGB & 137.460G & 0.0065 & 18.06 & 0.7743 & 17.10 & 0.7835 & 18.51 & 0.7977 \\
RIDNet \cite{anwar2019real} & short-exposure RGB & 439.714G & 0.0198 & 30.80 & 0.9095 & 31.03 & 0.9098 & 31.11 & 0.9091 \\
MIRNet \cite{zamir2020learning} & short-exposure RGB & 2.607T & 0.6893 & 31.67 & \textcolor{green}{0.9193} & 31.89 & 0.9219 & 32.03 & 0.9235 \\
MPRNet \cite{zamir2021multi} & short-exposure RGB & 2.642T & 0.1516 & 30.44 & 0.9043 & 31.24 & 0.9127 & 31.32 & 0.9136 \\
HINet \cite{chen2021hinet} & short-exposure RGB & 769.487G & 0.0313 & 30.32 & 0.8903 & 27.24 & 0.8253 & 28.78 & 0.8791 \\
\hline
QRNet & dual-exposure Quad-Bayer & \textcolor{red}{34.628G} & 0.0178 & \textcolor{red}{33.24} & \textcolor{red}{0.9403} & \textcolor{red}{33.61} & \textcolor{red}{0.9428} & \textcolor{red}{33.81} & \textcolor{red}{0.9445} \\
\hline
\end{tabular}
\end{center}
\end{table*}

\begin{figure*}[t]
\centering
\includegraphics[width=0.95\linewidth]{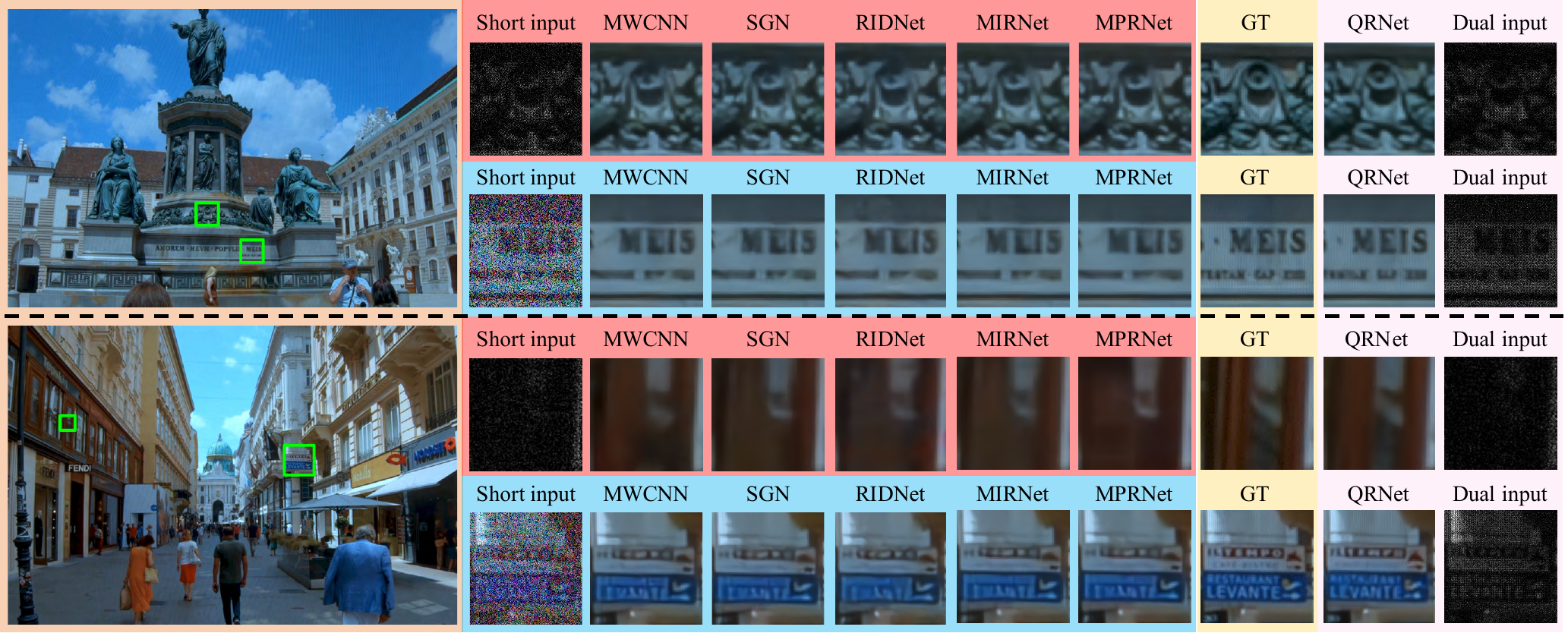}
\centering
\caption{Illustration of image restoration results of QRNet and other methods based on short-exposure images. The noise parameter is $K=0.25, \sigma=5$. The left is full-resolution ground truth images. On the right, input short-exposure Quad-Bayer image patches and generated image patches are in the red background; corresponding input short-exposure RGB image patches and generated image patches are in the blue background; corresponding input dual-exposure Quad-Bayer image patches and generated image patches are in the pink background. Ground truth image patches are in the yellow background.}
\label{fig_com1}
\end{figure*}

\begin{table*}[t]
\begin{center}
\caption{The comparisons of QRNet and other baselines based on long-exposure images.}
\label{table_com2}
\begin{tabular}{l|c|c|c|cc|cc|cc}
\hline
\multirow{2}{*}{Method} & \multirow{2}{*}{Input data structure} & MACs & Running time & \multicolumn{2}{c|}{$K=0.25$, $\sigma=5$} & \multicolumn{2}{c|}{$K=0.5$, $\sigma=5$} & \multicolumn{2}{c}{$K=0.75$, $\sigma=5$} \cr & & (512$\times$512) & (sec / 512$\times$512) & PSNR & SSIM & PSNR & SSIM & PSNR & SSIM \\
\hline
DeepDeblur \cite{nah2017deep} & long-exposure Quad-Bayer & 1.398T & 0.0366 & 26.37 & 0.8685 & 26.66 & 0.8757 & 25.30 & 0.8243 \\
SRN \cite{tao2018scale} & long-exposure Quad-Bayer & 343.004G & 0.0545 & 27.56 & 0.8988 & 27.66 & 0.9007 & 27.72 & 0.9027 \\
DeblurGAN \cite{kupyn2018deblurgan} & long-exposure Quad-Bayer & 205.445G & 0.0111 & 27.40 & 0.8967 & 27.51 & 0.8990 & 27.55 & 0.9001 \\
DeblurGANv2 \cite{kupyn2019deblurgan} & long-exposure Quad-Bayer & \textcolor{blue}{52.730G} & 0.0249 & 21.66 & 0.8439 & 21.08 & 0.8438 & 21.45 & 0.8474 \\
DMPHN \cite{zhang2019deep} & long-exposure Quad-Bayer & \textcolor{green}{202.458G} & 0.0995 & 27.50 & 0.8982 & 27.67 & 0.9010 & 27.68 & 0.9019 \\
MPRNet \cite{zamir2021multi} & long-exposure Quad-Bayer & 2.642T & 0.1435 & 27.07 & 0.8915 & 27.20 & 0.8954 & 27.19 & 0.8962 \\
HINet \cite{chen2021hinet} & long-exposure Quad-Bayer & 769.487G & 0.0315 & 27.08 & 0.8880 & 25.89 & 0.8143 & 25.08 & 0.8022 \\
MIMOUNet \cite{cho2021rethinking} & long-exposure Quad-Bayer & 300.205G & 0.0394 & 27.60 & 0.9007 & 27.73 & 0.9030 & 27.79 & 0.9041 \\
MIMOUNet++ \cite{cho2021rethinking} & long-exposure Quad-Bayer & 687.105G & 0.0831 & 27.64 & \textcolor{green}{0.9014} & 27.76 & 0.9034 & 27.81 & 0.9048 \\
QRNet & long-exposure Quad-Bayer & \textcolor{red}{34.628G} & 0.0178 & \textcolor{blue}{27.78} & \textcolor{blue}{0.9031} & \textcolor{blue}{27.95} & \textcolor{blue}{0.9058} & \textcolor{blue}{27.94} & \textcolor{blue}{0.9062}  \\
\hline
DeepDeblur \cite{nah2017deep} & long-exposure RGB & 1.398T & 0.0365 & 27.32 & 0.8932 & 27.06 & 0.8857 & 26.93 & 0.8827 \\
SRN \cite{tao2018scale} & long-exposure RGB & 343.004G & 0.0563 & \textcolor{green}{27.68} & 0.9006 & \textcolor{green}{27.82} & \textcolor{green}{0.9029} & \textcolor{green}{27.86} & \textcolor{green}{0.9037} \\
DeblurGAN \cite{kupyn2018deblurgan} & long-exposure RGB & 207.123G & 0.0092 & 27.39 & 0.8966 & 27.44 & 0.8982 & 27.54 & 0.8996 \\
DeblurGANv2 \cite{kupyn2019deblurgan} & long-exposure RGB & \textcolor{blue}{54.408G} & 0.0258 & 20.51 & 0.8308 & 20.64 & 0.8350 & 20.51 & 0.8368 \\
DMPHN \cite{zhang2019deep} & long-exposure RGB & \textcolor{green}{202.458G} & 0.0998 & 17.04 & 0.7968 & 16.06 & 0.7604 & 17.88 & 0.8006 \\
MPRNet \cite{zamir2021multi} & long-exposure RGB & 2.642T & 0.1516 & 27.34 & 0.8928 & 27.49 & 0.8959 & 27.57 & 0.8981 \\
HINet \cite{chen2021hinet} & long-exposure RGB & 769.487G & 0.0313 & 21.53 & 0.8132 & 23.60 & 0.8396 & 22.20 & 0.8169 \\
MIMOUNet \cite{cho2021rethinking} & long-exposure RGB & 300.205G & 0.0397 & 27.60 & 0.9008 & 27.67 & 0.9024 & 27.78 & 0.9037 \\
MIMOUNet++ \cite{cho2021rethinking} & long-exposure RGB & 687.105G & 0.0835 & 27.64 & 0.9013 & 27.71 & 0.9025 & 27.74 & 0.9032 \\
\hline
QRNet & dual-exposure Quad-Bayer & \textcolor{red}{34.628G} & 0.0178 & \textcolor{red}{33.24} & \textcolor{red}{0.9403} & \textcolor{red}{33.61} & \textcolor{red}{0.9428} & \textcolor{red}{33.81} & \textcolor{red}{0.9445} \\
\hline
\end{tabular}
\end{center}
\end{table*}

\begin{figure*}[t]
\centering
\includegraphics[width=0.95\linewidth]{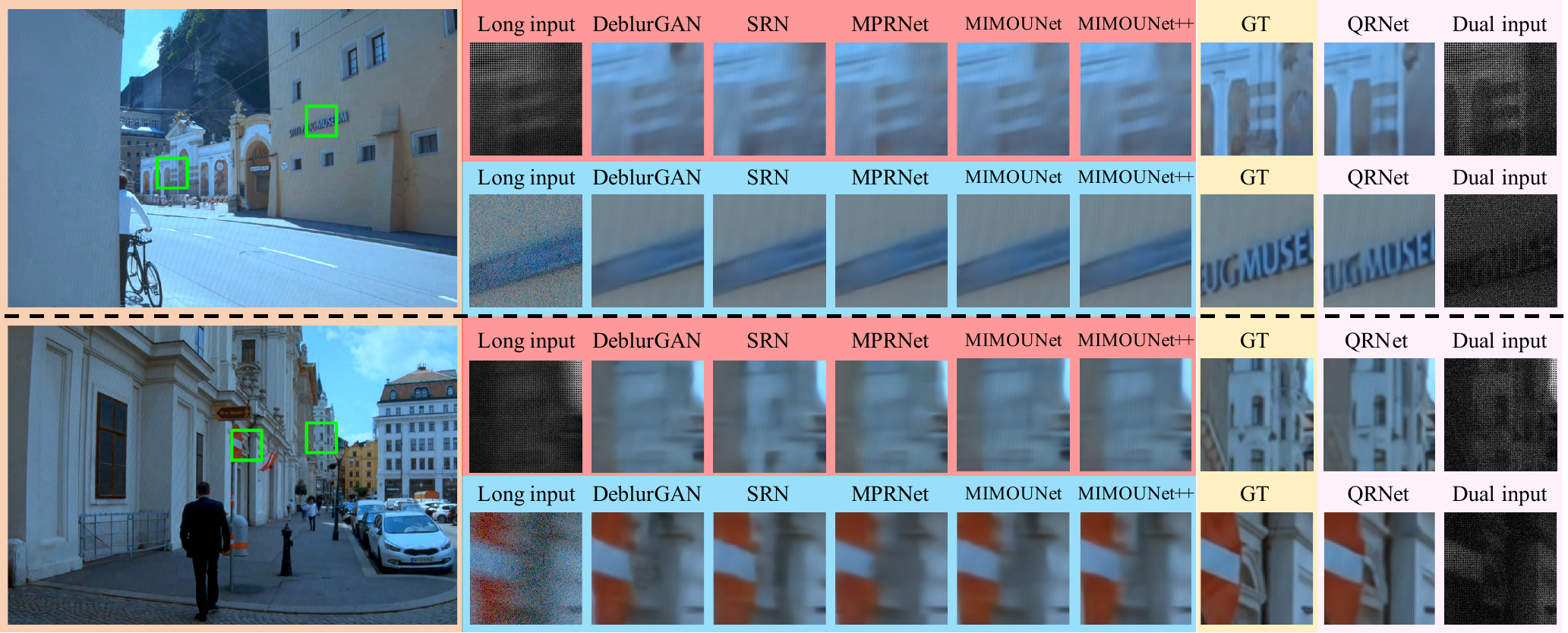}
\centering
\caption{Illustration of image restoration results of QRNet and other baselines based on long-exposure images. The noise parameter is $K=0.75, \sigma=5$. The left is full-resolution ground truth images. On the right, input long-exposure Quad-Bayer image patches and generated image patches are in the red background; corresponding input long-exposure RGB image patches and generated image patches are in the blue background; corresponding input dual-exposure Quad-Bayer image patches and generated image patches are in the pink background. Ground truth image patches are in the yellow background.}
\label{fig_com2}
\end{figure*}

\subsection{Evaluation with Short-exposure-based Methods}

In this section, we evaluate the QRNet based on the pipeline from dual-exposure Quad-Bayer images to clean and sharp RGB $y$ with methods based on short-exposure Quad-Bayer images or RGB images. The methods include IRCNN \cite{zhang2017learning}, DnCNN \cite{zhang2017beyond}, MemNet \cite{tai2017memnet}, MWCNN \cite{liu2018multi}, SGN \cite{gu2019self}, DSWN \cite{liu2020densely}, ADNet \cite{tian2020attention}, RIDNet \cite{anwar2019real}, MIRNet \cite{zamir2020learning}, MPRNet \cite{zamir2021multi}, and HINet \cite{chen2021hinet}.

The quantitative analysis is concluded in Table \ref{table_com1}. From the results, we draw four findings. First, the dual-exposure Quad-Bayer is superior for image denoising than only short exposure since it also contains long-exposure pixels, which are less noisy. Neural networks can utilize such information to improve the details and textures. For instance, SGN obtains a PSNR of 32.00 for dual exposure, while it is 31.78 (-0.22) for short exposure when $K=0.25, \sigma=5$. Secondly, Quad-Bayer is better for image restoration than RGB images since RAW data contains more information than the RGB counterpart. The top-performing method MIRNet \cite{zamir2020learning} based on short-exposure RGB obtains PSNR of 31.67, while SGN \cite{gu2019self} based on short-exposure Quad-Bayer obtains PSNR of 32.89 (+1.22), although it uses much fewer MACs than MIRNet. For SGN, it obtains PSNR of 31.78 when receiving Quad-Bayer while obtains 31.48 when receiving RGB images when $K=0.25, \sigma=5$. Thirdly, QRNet achieves better performance than existing methods by a large margin under different noise levels. It reveals that QRNet has good generalization ability. Finally, when changing the input data structure to the short-exposure Quad-Bayer for the QRNet, it still outperforms other methods. It further demonstrates that the improvement is induced by the QRNet network architecture.

We illustrate some restoration results in Figure \ref{fig_com1}. The results of QRNet are clearer and sharper than short-exposure-based methods. For instance, the letters ``MEIS'' in the second row and the letters ``LEVANTE'' in the fourth row of QRNet are much clearer than MWCNN, SGN, and MPRNet. This is because QRNet can use the information from long-exposure pixels, which are less noisy than short-exposure pixels. However, short-exposure-based methods can only exploit information from short-exposure pixels either in the Quad-Bayer RAW domain or RGB domain. Therefore, their performance is restricted when recovering the details and textures, while QRNet with dual-exposure Quad-Bayer can well address these problems.

\begin{figure*}[t]
\centering
\includegraphics[width=0.95\linewidth]{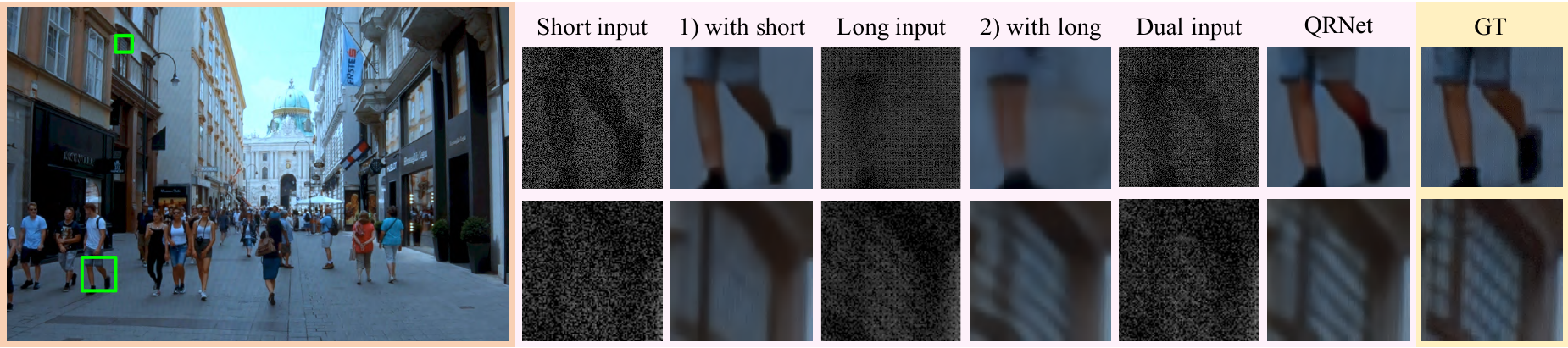}
\centering
\caption{Illustration of image restoration results of QRNet and changing input Quad-Bayer data structure ablation study settings. The noise parameter is $K=0.25, \sigma=5$. The left is full-resolution ground truth images. On the right, three pairs of input Quad-Bayer image patches and generated image patches (including ablation study settings and full QRNet) are in the pink background. Ground truth image patches are in the yellow background.}
\label{fig_ab1}
\end{figure*}

\begin{figure*}[t]
\centering
\includegraphics[width=0.95\linewidth]{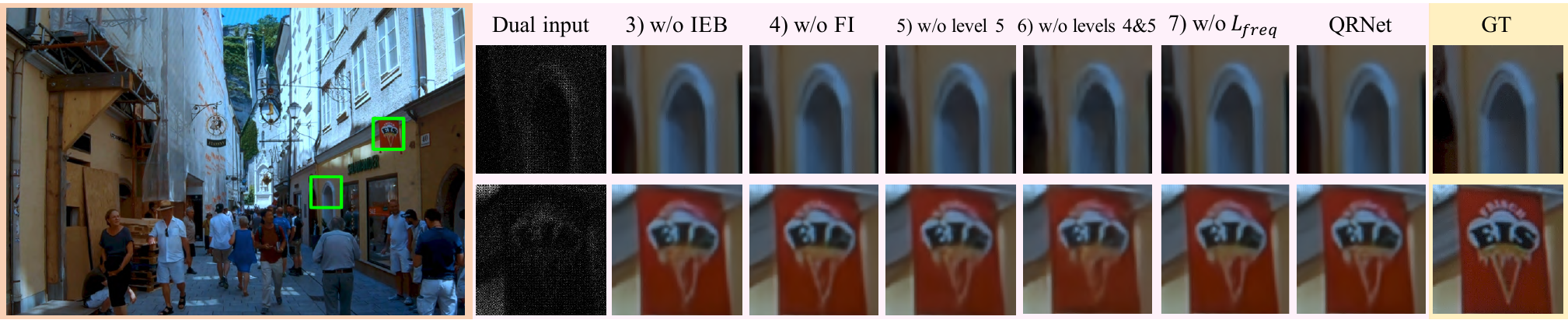}
\centering
\caption{Illustration of image restoration results of QRNet and other ablation study settings. The noise parameter is $K=0.25, \sigma=5$. The left is full-resolution ground truth images. In the right, input dual-exposure Quad-Bayer image patches and generated image patches (including ablation study settings and full QRNet) are in the pink background. Ground truth image patches are in the yellow background.}
\label{fig_ab2}
\end{figure*}

\subsection{Evaluation with Long-exposure-based Methods}

In this section, we evaluate the QRNet based on the pipeline from dual-exposure Quad-Bayer images to clean and sharp RGB $y$ with methods based on long-exposure Quad-Bayer images or RGB images. The methods include DeepDeblur \cite{nah2017deep}, SRN \cite{tao2018scale}, DeblurGAN \cite{kupyn2018deblurgan}, DeblurGANv2 \cite{kupyn2019deblurgan}, DPMHN \cite{zhang2019deep}, MPRNet \cite{zamir2021multi}, HINet \cite{chen2021hinet}, MIMOUNet \cite{cho2021rethinking}, and MIMOUNet++ \cite{cho2021rethinking}.

The quantitative analysis is concluded in Table \ref{table_com2}. On one hand, the proposed QRNet performs better than other long-exposure-based methods. Since the short-exposure pixels are captured in the dual-exposure Quad-Bayer, they can help deblurring due to no motion blur, although they include more noises. QRNet can well utilize this information to estimate the motion blur. Moreover, QRNet still outperforms other methods when receiving long-exposure Quad-Bayer input, e.g., QRNet obtains PSNR of 27.78 while MIMOUNet++ obtains 27.64 when $K=0.25, \sigma=5$. On the other hand, the state-of-the-art network architecture MIMOUNet++ trained on dual-exposure Quad-Bayer is still inferior to QRNet, even under different noise parameters, as concluded in Table \ref{table_sota2}. It demonstrates that the QRNet architecture is more appropriate for restoring dual-exposure Quad-Bayer data.

The qualitative analysis is illustrated in Figure \ref{fig_com2}. When there is motion blur in the captured images, which is common in real-world scenarios, QRNet is robust enough to recover sharp and clean images. However, the image deblurring methods based on either single-exposure Quad-Bayer RAW or RGB images cannot handle large blurs.

\subsection{Ablation Study}

To demonstrate the effectiveness of several components of QRNet, we define some ablation study settings:

\begin{enumerate}\setlength{\itemsep}{-0.0cm}
\item with short: replacing dual-exposure Quad-Bayer input with short-exposure Quad-Bayer input;

\item with long: replacing dual-exposure Quad-Bayer input with long-exposure Quad-Bayer input;

\item w/o IEB: replacing input enhancement block (IEB) with a convolutional layer receiving only Pixel Unshuffle downsampled data;

\item w/o FI: dropping all the feature interactions of levels 3 to 5;

\item w/o level 5: dropping all layers of level 5;

\item w/o level 4\&5: dropping all layers of levels 4 and 5;

\item w/o $L_{freq}$: dropping the frequency loss $L_{freq}$ ($\alpha = 0$).

\end{enumerate}

\begin{table}[t]
\begin{center}
\caption{The ablation study comparisons of QRNet.}
\label{table_ab}
\begin{tabular}{l|c|c|cc}
\hline
\multirow{2}{*}{Setting} & \multirow{2}{*}{Description} & MACs & \multicolumn{2}{c}{$K=0.25$, $\sigma=5$} \cr & & (512$\times$512) & PSNR & SSIM \\
\hline
with short & Input data & 34.628G & 32.89 & 0.9372 \\
with long & Input data & 34.628G & 27.78 & 0.9031 \\
\hline
w/o IEB & Network & 33.084G & 33.15 & 0.9391 \\
w/o FI & Network & 31.944G & 33.14 & 0.9391 \\
w/o level 5 & Network & 21.539G & 32.79 & 0.9349 \\
w/o level 4\&5 & Network & 12.475G & 31.97 & 0.9238 \\
\hline
w/o $L_{freq}$ & Loss function & 34.628G & 33.13 & 0.9389 \\
\hline
QRNet & Full QRNet & 34.628G & \textcolor{red}{33.24} & \textcolor{red}{0.9403} \\
\hline
\end{tabular}
\end{center}
\end{table}

We conduct the experiments on the QR dataset validation set. The quantitative results are concluded in Table \ref{table_ab} and some results are illustrated in Figure \ref{fig_ab1} and \ref{fig_ab2}.

\noindent \textbf{Input data.} Dual-exposure Quad-Bayer data is the most significant for QRNet to obtain the performance gain. Compared with only using short- or long-exposure data, dual exposure brings 0.35dB and 5.46dB PSNR gains. It shows that dual-exposure Quad-Bayer is superior to single exposure since multiple exposures can provide more information. As shown in Figure \ref{fig_ab1}, QRNet with dual-exposure Quad-Bayer produces sharper outputs than setting 1), e.g., the window. As the long-exposure pixels are less noisy, QRNet can extract the texture and detail information from them. The QRNet outputs are also less blurry than setting 2), e.g., the leg. Since the short-exposure pixels provide accurate position information, QRNet can better estimate the motion field with dual-exposure data than only using long-exposure pixels.

\noindent \textbf{Network.} All the proposed input enhancement block (IEB), feature interaction (FI), and multi-level feature extraction are helpful in improving image restoration quality. For instance, settings 5) and 6) generate a deformed archdoor (first row); and settings 3-6) produce blurry flags (second row), as shown in Figure \ref{fig_ab2}. However, the results full QRNet have more distinguishable and sharper edges. From Table \ref{table_ab}, we can see IEB, FI, level 5 of QRNet, and level 4\&5 of QRNet bring 0.09dB, 0.1dB, 0.45dB, and 1.27dB PSNR gains. They demonstrate every network component is helpful in recovering textures and details.

\noindent \textbf{Loss function.} From setting 7), we can see dropping $L_{freq}$ leads to decreases of PSNR (-0.11dB) and SSIM metrics. Also as shown in Figure \ref{fig_ab2}, the edges at the top of the flag of setting 7) are blurrier than the full QRNet. It demonstrates that frequency loss can enhance detail recovery.

In conclusion, the dual-exposure data structure, network components, and loss function are beneficial to improving the image restoration quality for QRNet.

\section{Conclusion}

In this paper, we have presented a novel pipeline for joint image deblurring and denoising of Quad-Bayer RAW images. The Quad-Bayer pattern provides complementary exposures that enable both noise reduction and motion deblurring. To effectively restore clean and sharp RGB images from Quad-Bayer inputs, we have proposed QRNet, an efficient hierarchical neural network architecture. A key component of QRNet is the input enhancement block which reduces artifacts from downsampling and enhances details. Inter-level feature interactions further improve feature representations by sharing information between network levels. We have also introduced a new Quad-Bayer to RGB mapping dataset (QR dataset) to train and benchmark QRNet. Experiments demonstrate state-of-the-art performance in joint deblurring and denoising using our approach. The proposed QRNet achieves superior image restoration quality compared to existing methods while using less computational costs.

\section*{Acknowledgment}

The authors would like to thank the anonymous reviewers and editors for many helpful comments. The authors would also like to thank the computational imaging group of SenseTime Research and Tetras.AI for hardware support and helpful discussions.

\ifCLASSOPTIONcaptionsoff
  \newpage
\fi



%

{
\bibliographystyle{IEEEtran}
\bibliography{IEEEabrv,mybibfile}
}

%
%

%

\begin{IEEEbiography}[{\includegraphics[width=1in,height=1.25in,clip,keepaspectratio]{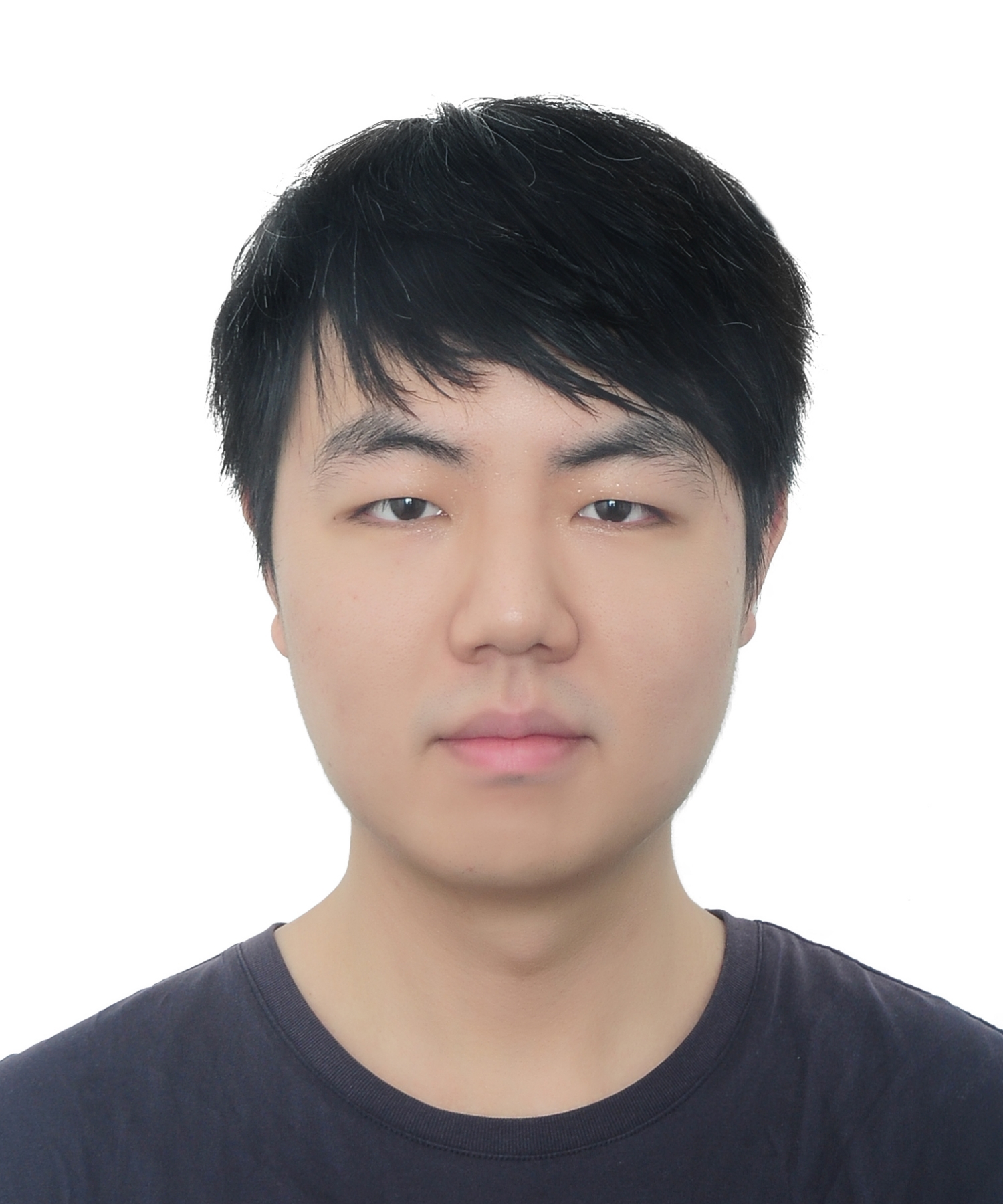}}]{Yuzhi Zhao}

(S'19) received the B.Eng. degree in Electronic and Information Engineering from Huazhong University of Science and Technology (HUST) in 2018, and the Ph.D. degree in Electronic Engineering from City University of Hong Kong (CityU) in 2023.

His research interests include low-level vision, computational photography, and generative models (such as diffusion models and multimodal large language models). He has authored 10 first-authored works and has contributed to over 35 papers in prestigious venues. He also serves as a peer reviewer for international conferences and journals, including CVPR, ICCV, ECCV, SIGGRAPH, AAAI, PR, IJCV, CVIU, IEEE TMM, IEEE TCSVT, IEEE TNNLS, IEEE TIP, ACM TOG, etc.

\end{IEEEbiography}

\begin{IEEEbiography}[{\includegraphics[width=1in,height=1.25in,clip,keepaspectratio]{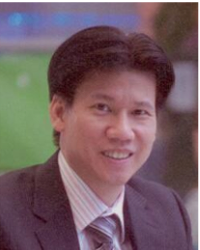}}]{Lai-Man Po}

(M’92–SM’09) received the B.S. and Ph.D. degrees in electronic engineering from the City University of Hong Kong, Hong Kong, in 1988 and 1991, respectively. He has been with the Department of Electronic Engineering, City University of Hong Kong, since 1991, where he is currently an Associate Professor of Department of Electrical Engineering. He has authored over 150 technical journal and conference papers. His research interests include image and video coding with an emphasis deep learning based computer vision algorithms.

Dr. Po is a member of the Technical Committee on Multimedia Systems and Applications and the IEEE Circuits and Systems Society. He was the Chairperson of the IEEE Signal Processing Hong Kong Chapter in 2012 and 2013. He was an Associate Editor of HKIE Transactions in 2011 to 2013. He also served on the Organizing Committee, of the IEEE International Conference on Acoustics, Speech and Signal Processing in 2003, and the IEEE International Conference on Image Processing in 2010.

\end{IEEEbiography}

\begin{IEEEbiography}[{\includegraphics[width=1in,height=1.25in,clip,keepaspectratio]{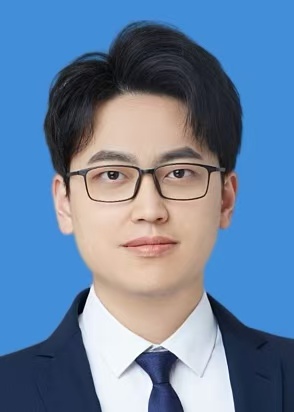}}]{Xin Ye}

received the B.S. degree in physics from Zhejiang University in 2014 and received the Ph.D. degree in physics from the Chinese University of Hong Kong in 2019. His research interests include image processing and deep learning.

\end{IEEEbiography}

\begin{IEEEbiography}[{\includegraphics[width=1in,height=1.25in,clip,keepaspectratio]{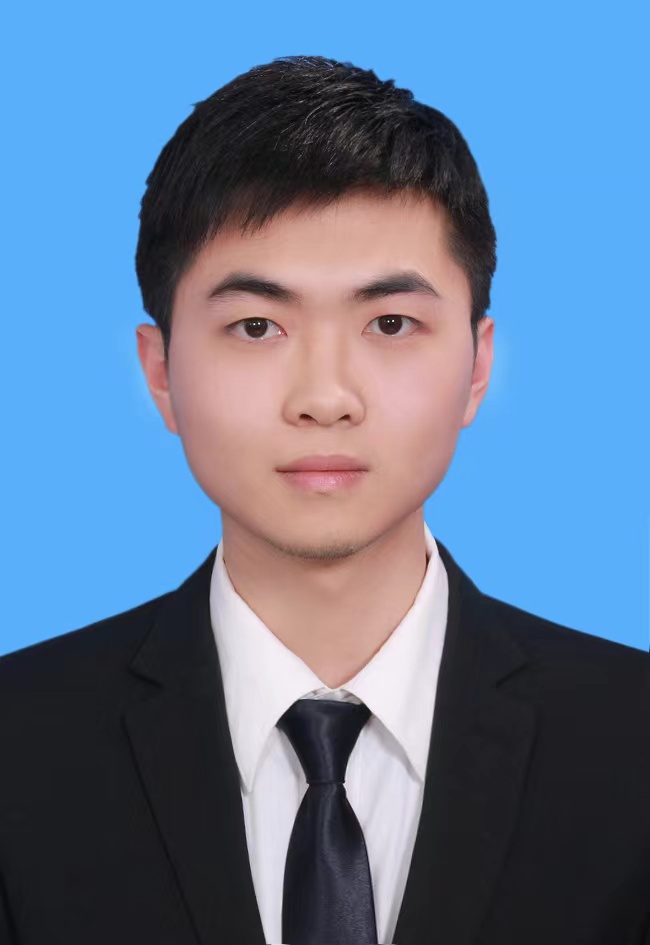}}]{Yongzhe Xu}

received the B.Eng. and M.Eng. degrees in software engineering from Sun Yat-sen University, Guangzhou, China, in 2017 and 2019, respectively. He is currently an algorithm engineer with the Computational Imaging Group at Tetras.ai. His research interests include generative models, computational photography , and deep learning.

\end{IEEEbiography}

\begin{IEEEbiography}[{\includegraphics[width=1in,height=1.25in,clip,keepaspectratio]{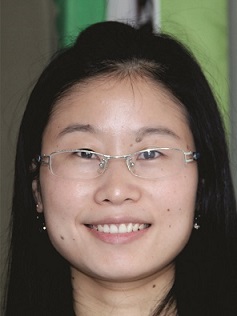}}]{Qiong Yan}

received her Ph.D. degree in computer science and engineering from Chinese University of Hong Kong in 2013 and the Bachelor’s degree in computer science and technology from University of Science and Technology of China in 2009. She is now a research director in SenseTime, leading a group on computational imaging related research and production. Her research focuses on low-level vision tasks, such as image/video restoration and enhancement, image editing and generation.

\end{IEEEbiography}

%




\end{document}